\documentclass[11pt,a4paper]{article}
\usepackage{jheppub}

\usepackage{mathtools,tabularx}
\usepackage{graphicx,amsfonts,color,comment,amsmath,float}
\usepackage{amssymb}
\usepackage{mathrsfs,amssymb}

\newcommand{\eps}{\epsilon}

\DeclareMathOperator{\diag}{diag}

\newcommand{\be}{\begin{equation}}
\newcommand{\ee}{\end{equation}}

\begin{document}

\title{Activating the 4th Neutrino of the $3+1$ Scheme }

\author[a,b]{Peter B.~Denton,}

\author[c]{Yasaman Farzan,}

\author[d]{and Ian M.~Shoemaker}

\affiliation[a]{Niels Bohr International Academy, University of Copenhagen, The Niels Bohr Institute,
Blegdamsvej 17, DK-2100, Copenhagen, Denmark}
\affiliation[b]{Physics Department, Brookhaven National Laboratory, Upton, New York 11973, USA}
\affiliation[c]{School of physics, Institute for Research in Fundamental Sciences (IPM)
P.O.Box 19395-5531, Tehran, Iran}
\affiliation[d]{Department of Physics, University of South Dakota, Vermillion, SD 57069, USA}
\emailAdd{peterbd1@gmail.com}
\emailAdd{yasaman@theory.ipm.ac.ir}
\emailAdd{ian.shoemaker@usd.edu}

\date{\today}

\abstract{
Non-Standard Interactions (NSI) of neutrinos with matter has received renewed interest in recent years. In particular, it has been shown that NSI can reconcile the $3+1$ solution with IceCube atmospheric data with $E_\nu >500$ GeV, provided that the effective coupling of NSI is large, e.g. $\sim 6 G_F$. The main goal of the present paper is to show that contrary to intuition, it is possible to build viable models with large NSI by invoking a new $U(1)$ gauge symmetry with gauge boson of mass $\sim 10$ eV. We refer to these new constructions as $3+1+ U(1)$ models.  In the framework of a $3+1$ solution to LSND and MiniBooNE anomalies, we show that this novel NSI can help to solve the tension with cosmological bounds and constraints from IceCube atmospheric data with $E_\nu>500$ GeV.	
We then discuss the implications of the MINOS and MINOS+ results for the  3+1+$U(1)$ scenario.
}


\maketitle
\newpage
\section{Introduction}
Adding a light sterile neutrino mixed with active neutrinos to the standard model is one of the most economic extensions which leads to a rich phenomenology. The longstanding LSND anomaly \cite{Aguilar:2001ty} has been largely confirmed by MiniBooNE \cite{Aguilar-Arevalo:2018gpe} recently reaching over 6$\sigma$ evidence by  combining the two data sets.
Along with the reactor \cite{Mention:2011rk} and Gallium \cite{Giunti:2010zu} anomalies there is a simple solution to these anomalies within the so-called 3+1 scheme which requires a sterile neutrino of mass $\sim 1$ eV with a mixing with active neutrinos, $\theta\sim \mathcal{O}(0.1)$. This scheme is however in serious tension with the observation of atmospheric neutrinos by IceCube and with cosmological constraints on the presence of new light neutrinos in the early universe.

Within the standard 3+1 scheme, the propagation in matter is governed by the following Hamiltonian
\be H = \frac{1}{2E}U\cdot \diag(0,\Delta m_{21}^2, \Delta m_{31}^2,\Delta m_{41}^2) \cdot U^\dagger+ \diag(V_e,V_\mu,V_\tau,V_s)\,, \ee
where $V_e =\sqrt{2} G_F N_e-(\sqrt{2}/2)G_FN_n$ and $V_\mu=V_\tau =-(\sqrt{2}/2)G_FN_n$ and $U$ is the $4\times 4$ unitary mixing matrix including the sterile neutrino. If sterile neutrinos have no interaction with matter, the corresponding matter potential vanishes, $V_s=0$.
For high energy neutrinos, $\Delta m_{41}^2/(2E)$ can be of order of $V_\mu$, leading to a resonant enhancement of the $\nu_\mu \to \nu_s$ oscillation which can be probed by atmospheric neutrinos at IceCube \cite{Nunokawa:2003ep,Choubey:2007ji,Razzaque:2011ab,Barger:2011rc,Esmaili:2012nz,Esmaili:2013cja,Lindner:2015iaa,Esmaili:2013vza}, see fig.~\ref{fig:probabilities}.
Null results from IceCube on the deviation of $P(\nu_\mu \to \nu_\mu)$ from the standard 3$\nu$ scheme prediction set a bound on active sterile mixing which is in tension with the value derived from LSND \cite{TheIceCube:2016oqi}. Ref. \cite{Liao:2016reh} suggested  turning on NSI for $\nu_\mu$ and $\nu_\tau$ to suppress the effective mixing for $E_\nu>500$ GeV.  However, as shown in \cite{Liao:2016reh}, this requires values of NSI which are larger than the standard weak coupling. It is very challenging to build viable models with such large NSI couplings that satisfy various bounds on the couplings of neutrinos and active neutrinos. However, the bounds on the coupling of sterile neutrinos are more relaxed so it is intriguing to entertain the possibility of a fourth neutrino with sizeable interaction with matter fields even if MiniBooNE and LSND anomalies are refuted by the upcoming SNB experiment.

 Let us suppose the sterile neutrino (in the flavor basis) has an effective interaction of the following form with matter fields ($f=u,d$)
\be \mathscr{L}=-2\sqrt{2}~ \epsilon ~G_F ~(\bar{\nu}_s \gamma^\mu P_L \nu_s) (\bar{f}\gamma_\mu f)\,,\label{s-bar}\ee
where $P_L\equiv (1-\gamma_5)/2$ is the left-handed projector.

The value of the sterile matter potential in Earth can be then estimated as  \be V_s= \label{Vs}2.5 \times 10^{-12}~{\rm eV} \left(\frac{\rho}{5~{\rm gr~cm}^{-3}}\right) \epsilon\,,\ee which should be compared to $10^{-12}$ eV $(\Delta m^2/$eV$^2)($TeV$/E)$.
The active sterile mixing in Earth will be suppressed  for neutrinos with energy larger than TeV if $\epsilon \gtrsim 10$. On the other hand, to reproduce LSND and MiniBooNE we need unsuppressed mixing for energies smaller than few hundred MeV so we obtain an upper bound on $\epsilon<10^4$.
Intermediate values of $\epsilon$ can dramatically affect the long baseline and atmospheric neutrino data in the energy range $10~{\rm GeV}<E<100~{\rm GeV}$ which has been detected by IceCube. Thus, to solve the tension, we will focus on $|\epsilon| \lesssim 10$.
According to the MiniBooNE collaboration, the data can be better explained by an agnostic introduction of ``new" matter effects with a resonance energy of 300 MeV\footnote{See slide 19, E-C Huang, Neutrino 2018 \cite{huang_en_chuan_2018_1287004}.}. We shall comment on this possibility within the framework that we are discussing.

Another tension shows up between the $3+1$ solution to the short baseline neutrino observation and cosmological bounds.
It has been shown in a series of papers \cite{Capozzi:2017auw,Hannestad:2013ana,Dasgupta:2013zpn,Mirizzi:2014ama,Cherry:2014xra,Chu:2015ipa,Cherry:2016jol,Vecchi:2016lty,Saviano:2014esa,Ade:2015xua,Song:2018zyl}
that self-interaction of sterile neutrinos can ease this tension. We shall discuss under what conditions the interactions that we are discussing can solve the tension.
Long baseline NOvA \cite{Adamson:2017zcg} and MINOS and MINOS+ \cite{Adamson:2017uda} experiments can also constrain the 3+1 scenario. In fact, according to \cite{Adamson:2017uda}, the constraints from MINOS+ can rule out a significant part of the 3+1 solution to LSND. However, the strength of these bounds are debated in \cite{Louis:2018yeg}.  We  study the bounds that MINOS+ can set on various combinations of $\theta_{14}$, $\theta_{34}$, and $\theta_{24}$ with and without interaction of sterile neutrinos with matter.  

The paper is organized as follows. 
In section \ref{bound}, we enumerate the various bounds that already exist on the new couplings and show that despite these bounds, it is  still possible to obtain $\epsilon \gtrsim1$. In section \ref{model}, we describe a  model that can lead to the couplings we are interested in. We show that the region of the parameter space that leads to $\epsilon\gtrsim1$ is natural in the sense that it does not suffer from fine tuning. In section \ref{osc}, we first quantify how turning on $\epsilon$ can reconcile TeV range atmospheric neutrino data with the 3+1 solution to the LSND. We then study the bounds from MINOS and MINOS+. In section \ref{summary}, we summarize our results.

\section{Bounds on new gauge couplings of neutrinos and baryons \label{bound}}
\begin{figure}
\centering
\includegraphics[width=0.6\textwidth]{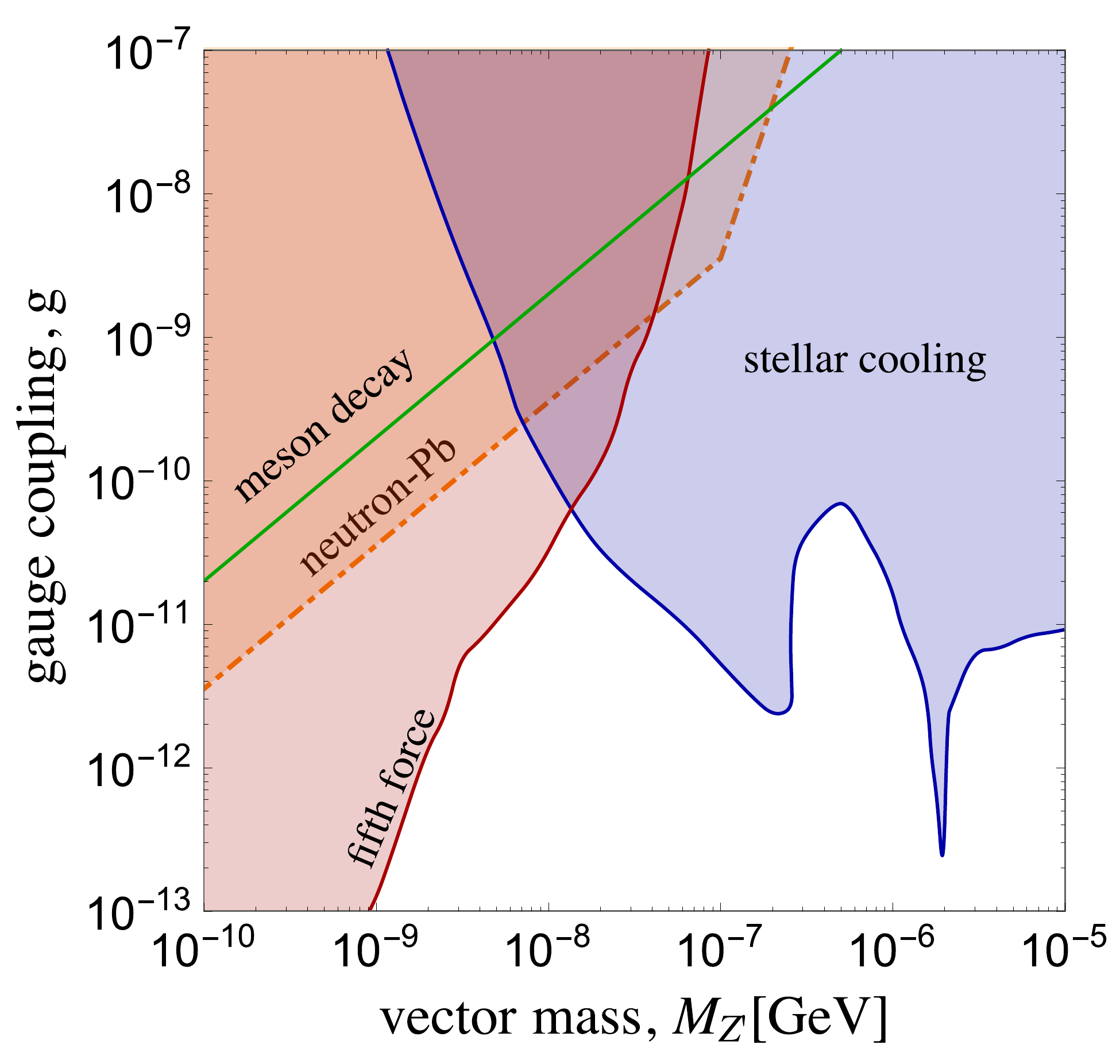}
\caption{Here we display some of the most relevant bounds on the $B-L$ gauge coupling in the mediator mass range of interest. These include stellar cooling in the Sun and horizontal branch stars in globular clusters~\cite{Redondo:2013lna,Hardy:2016kme}, bounds from  $n-$Pb scattering data \cite{Leeb:1992qf}, meson decays~\cite{Artamonov:2009sz,Batell:2014yra}, and fifth force searches~\cite{Bordag:2001qi}.  }
\label{fig:Bounds}
\end{figure}
In this section, we outline the various bounds that constrain the couplings of $Z^\prime$ to baryons ($g_B$) and to a sterile neutrino $\nu_s$ ($g_s$). Let us start with the bounds on $g_B$.
These bounds are shown in Fig.~\ref{fig:Bounds}.

{\it Bound from fifth force:} For $m_{Z^\prime}<10$ eV, the most stringent bound comes from the so-called 5th force search experiments.  By introducing this new interaction, two test objects located at a distance of $r$ from each other will  exert a new force to one another with a potential given by
\be g_B^2 \frac{\exp[{-m_{Z^\prime} r}]}{r} \frac{m_1m_2}{m_p^2}\left(1+a_e \frac{Z_1}{A_1} \right) \left(1+a_e \frac{Z_2}{A_2} \right) \ee
where  $m_i$, $A_i$ and $Z_i$ (with $i = 1,2$) are respectively the mass, mass number and atomic number of the test objects. Here, $a_e$ represents the ratio of the coupling to the electron to $g_B$. For example,  in case of $B-L$ gauge symmetry, $a_e=-1$.  Since the test objects are electrically neutral, $Z_i$ gives the numbers of protons as well as that of electrons in atoms of test objects. For $m_{Z^\prime} \sim 10$~eV  ({\it i.e.,} range of $2\times 10^{-6}$ cm), the measurement of the Casimir force (see fig.~27 of \cite{Bordag:2001qi}) sets a strong bound  on $g_B$ as shown in fig.~\ref{fig:Bounds} with a red line, marked with fifth force.

{\it Stellar cooling:} Another strong bound on the model comes from the cooling rate of the Sun and Horizontal Branch (HB) stars. The $Z^\prime$ boson can be produced in the stellar core via plasmon effect.  For $m_{Z^\prime}\sim 100 $ eV (for $m_{Z^\prime}\sim$ keV) the production of both transverse and longitudinal  $Z^\prime$ in the sun (in HB stars) can be on-shell. Thus, for this mass range stellar cooling can set a very strong bound on couplings to matter fields. For smaller $m_{Z^\prime}$, the on-shell production of the transverse $Z^\prime$ is suppressed but longitudinal $Z^\prime$ can still be produced on-shell contributing to the cooling of the star. Considering the upper bound on the cooling rate, Refs.~\cite{Redondo:2013lna,Hardy:2016kme} find strong upper bound on the $Z^\prime$ coupling to electrons from $Z^\prime$ production in stars. This bound is shown with a blue line in fig.~\ref{fig:Bounds} which corresponds to $|a_e|=1$ ({\it e.g.,} for gauged $B-L$).

{\it Bounds from neutron scattering and meson decay:} Another bound comes from the $n-$Pb scattering experiment \cite{Leeb:1992qf} which is shown in fig.~\ref{fig:Bounds} with dashed-dotted orange line.
For $m_{Z^\prime}>30$ MeV, the strongest bound comes from  $K^+ \to \pi^++{\rm missing ~energy}$ \cite{Artamonov:2009sz,Batell:2014yra} but this bound is irrelevant for $m_{Z'}<$keV. Notice that in our model, $Z^\prime$ decays to $\nu_s \bar{\nu}_s$ appearing as missing energy so the recent searches for dark photon from NA48/2 which looks
for $\pi^0 \to Z^\prime \gamma$, $Z^\prime \to e^-e^+$ are not relevant for our model.
Moreover, the NOMAD bound on
$\pi^0 \to Z^\prime \gamma$ \cite{Gninenko:1998pm,Altegoer:1998qta} comes from search for subsequent production of $\pi^0$ from $Z^\prime$ scattering off nuclei in detector.
The NOMAD bound does not apply to our case because in our model, $Z^\prime$ decays to $\nu_s\bar{\nu}_s$ before reaching the detector.
From the three-body decay of charged mesons ($K^+ ({\it or}~\pi^+) \to \nu Z' \mu^+( {\it or}~e^+)$), strong bounds on  the coupling of $Z'$
to active neutrinos ($g_{\alpha \beta } \bar{\nu}_\alpha \gamma^\mu \nu_\beta Z'_\mu$)  can be derived $\sqrt{\sum_\alpha|g_{e \alpha}|^2},\sqrt{\sum_\alpha|g_{\mu \alpha}|^2}\lesssim 2 \times 10^{-9} (m_{Z'}/10~{\rm eV})$ \cite{Bakhti:2017jhm,Laha:2013xua,Bakhti:2018avv} which in the context of $B-L$ models also applies to $g_B$ and is shown in fig.~\ref{fig:Bounds}.

Let us now discuss the bound on $g_s$. Since no sterile neutrino has so far been discovered, it is no surprise that the bounds on $g_s$ are not very strong. However, from cosmological observations, the following constraints can be already set on $g_s$.

{\it Neutrino decay:} Because of the neutrino mixing, the coupling of $g_s$ to $\nu_s$ can lead to decay of the heavier neutrino mass eigenstates to lighter ones. If $Z'$ is heavier than $\nu_4$, the two-body decay will be forbidden but the three-body decay $\nu_i \to \nu_j
\bar\nu_k \nu_l$ can take place where $\nu_i$ can be any neutrino mass eigenstate other than the lightest one. The decay rate of $\nu_i$ with energy $E_\nu$ can be estimated as
\[\Gamma_i = \frac{g_s^4|U_{si}U_{sj}U_{sk}U_{sl}|^2}{192 \pi^3} \frac{m_i^5}{m_{Z'}^4}\frac{m_i}{E_\nu}\]
where the mass of the final neutrinos are neglected. The factor $m_i/E_\nu$ takes care of the time dilation because neutrinos decay in flight. Taking $U_{s4}\sim 1$, $U_{s i}|_{i \in \{ 1,2,3\}} \sim 0.1$, $m_{Z'}\sim 10$~eV and $m_4 \sim 1$ eV, we find $c \tau_4 =c \Gamma_4^{-1}\sim 5 \times 10^{31} ~
{\rm cm} (3\times 10^{-4}/g_s)^4 (E_\nu/300~{\rm MeV})$. This means that $\nu_4$ produced in the oscillation of the atmospheric neutrinos,
in long (and, of course, short) baseline neutrino beams or in the way from the Sun cannot decay before reaching the detector. However,
 $\nu_4$ produced in the oscillation of cosmic neutrinos can decay. Moreover, as long as $g_s \gtrsim3 \times 10^{-4}$,  $\nu_4$ produced in the early universe can decay before recombination.
The decay of $\nu_i$ (where $i$ stands for 2 and 3 (1) for normal (inverted) ordering)  will be further suppressed by $(m_i^6 /m_4^6)U_{si}^2\ll 1$ so the decay cannot be relevant even for cosmic neutrinos.

 {\it Bounds from cosmology:}
We first discuss the bounds from $\sum m_\nu$ and free streaming of neutrinos at recombination. We then  address the production of $\nu_s$ before neutrino decoupling and the contribution from $\nu_s$ to extra-relativistic degrees of freedom,  $(\delta N_\nu)_{eff}$.

 As discussed above for $g_s \gtrsim3 \times 10^{-4}$,  $\nu_4$ decays into lighter neutrino states before the onset of structure formation so the bound on the sum of neutrino masses does not restrict our scenario (see also Ref.~\cite{Archidiacono:2015oma}).
We should however check the bounds on extra relativistic degrees of freedom and on free streaming of neutrinos.

 The self-interactions of $\nu_s$
 induce a nonzero $V_s$ in early universe which suppresses the active sterile mixing and therefore the sterile
neutrino production before neutrino decoupling and BBN era. The bounds from BBN can be therefore relaxed if the $g_s$ coupling is larger than $O(10^{-3}-10^{-2})$  and the mass of the gauge boson coupled to $\nu_s$ is smaller than MeV \cite{Saviano:2014esa}.
The second condition can be readily fulfilled in our model. Let us check whether   $g_s\sim O(10^{-3})-O(10^{-2})$ can prevent neutrinos from free streaming  at structure formation.
As discussed before, for $g_s \gtrsim3 \times 10^{-4}$, the $\nu_4$ component of the active neutrinos decays away.
However, the $\nu_i$ component also has a coupling of $g_s U_{s i}^2\bar{\nu}_i\gamma^\mu \nu_i Z'_\mu$ which leads to a self-interaction cross section of
$g_s^4|U_{si}|^8T^2/(4\pi m_{Z'}^4)$ at $T^2\ll m_{Z'}^2$. For $g_s U_{si}^2>5 \times 10^{-6}$, this means that neutrinos will not be free steaming at the onset of structure formation so for $U_{si}\sim 0.1$, $g_s\sim \mathcal O(10^{-3})-\mathcal O(10^{-2})$ is ruled out by the free streaming of $\nu$  at recombination. The suppression of effective mixing at $T>$MeV therefore requires another mechanism which we will introduce later. Thus, we take $g_s\sim 3 \times 10^{-4}$ to guarantee the free streaming of all active neutrinos as well as fast decay of $\nu_4$.
Notice however that for $T>m_4$ process $\nu_i \bar{\nu}_i \to \nu_4
\bar{\nu}_4$ and $\nu_i\nu_i \to \nu_4\nu_4$ can take place with  mean free path shorter by a factor of $U_{si}^4$.
This sudden increase in the mean free path of $\nu_i$ in the threshold of structure formation may  leave an observable effect that requires more thorough investigation beyond the scope of the present paper.
For a fixed mixing angle, suppressing $\nu_i+\!\stackrel{(-)}{\nu_i} \to \nu_4+\!\stackrel{(-)}{\nu_4}$ requires smaller values of $g_s$ (less
than $5 \times 10^{-5}$) but then $\nu_4$ cannot decay to $\nu_i|_{i<4}$ fast enough. Introducing lighter degrees of freedom with relatively large coupling to $Z'$ can make fast enough decay of $\nu_4$ possible but we shall not explore this addition.

Within the scheme of the 3+1 solution to the LSND,  $\nu_s$ can be produced through oscillation in the early universe when $(\Delta m_{41}^2 /E_\nu)|_{E_\nu\sim T} (M_{Pl}^*/T^2)> 1$  or equivalently when
$$T<\sqrt[3]{\Delta m_{41}^2 M_{Pl}^*} \sim {\rm GeV}.$$
 The produced $\nu_s$ contribute to extra relativistic degrees of freedom on which there are strong bounds from
BBN and CMB. Let us see whether the new neutral current (NC) interaction between sterile neutrinos and quarks can induce an effective potential leading to suppression of effective neutrino mixing. As is well-known at high temperatures, because of charged current (CC) interactions between $\stackrel{(-)}{\nu}_e$ and $e^{\pm}$, an effective potential proportional to $G_F^2 T^4 E_\nu$ emerges which is non-zero even when the densities of $e^-$ and $e^+$ are equal. Since the new interaction that we turn on between sterile neutrinos and quarks is of neutral current type, we do not expect a similar effect here. The matter effects on $\nu_s$ need asymmetry between baryons and anti-baryons at $T<{\rm GeV}$. Since at these temperatures, baryons are non-relativistic, we can write $V_s=\sqrt{2}G_F\epsilon (3 n_B)$ where 3 reflects the fact that there are three valence quarks in each baryon. In order to have suppression of mixing due to NSI at temperature $T$, NSI should satisfy the bound
$$ \epsilon\gg \frac{(\Delta m_{41}^2/T)}{2\sqrt{2} G_F (3/2 n_B)}.$$
Inserting $n_B =\eta_B n_\gamma\sim 10^{-10} n_\gamma$ and $T\sim$ MeV, we find $\epsilon\gg 10^9$. Considering the strong bounds on the coupling of nuclei to new particles it seems impossible to have such large $\epsilon$.
Moreover, such large $\epsilon$ is ruled out by neutrino oscillation experiments \cite{Kopp:2014fha}.
However within asymmetric dark matter ($\chi$) scenario \cite{Kaplan:1991ah,Nussinov:1985xr,Barr:1991qn,Barr:1990ca,Gudnason:2006ug,Dodelson:1991iv,Fujii:2002aj,Kitano:2004sv,Kitano:2008tk,Farrar:2005zd,Kaplan:2009ag},  we may be able to achieve desired suppression of active sterile mixing  \cite{Dasgupta:2013zpn} invoking NSI of form
\be \label{s-chi} \mathcal{L}=-2\sqrt{2} \epsilon_\chi G_F (\bar{\nu}_s \gamma^\mu P_L \nu_s) (\bar{\chi}\gamma_\mu (1+b \gamma_5) \chi), \ee
where $b$ is an arbitrary real number.
For definiteness, let us take the benchmark point $n_\chi=n_B$ (and $m_\chi/m_p\simeq 5$) that is motivated by the scenario within which the dark matter and baryonic matter asymmetries are simultaneously created by the same mechanism. The suppression of active sterile mixing then requires
$$\epsilon_\chi\gg 10^9.$$

Such effective couplings shown in Eqs.~(\ref{s-bar},\ref{s-chi}) can originate from integrating out the intermediate $U(1)$ gauge boson, $Z^\prime$. Let us denote the coupling of $Z^\prime$ to quarks  with $g_B/3$ and  those to  $\nu_s$ and $\chi$  with respectively  $g_s$ and $g_\chi$.
{In general, $\chi$ and $\nu_s$ can have different  $U(1)$ charges.  A long as $g_\chi <10^{-4}$, the condition $m_\chi \alpha_\chi \ll m_{Z'}$ is satisfied so we are in the perturbative regime \cite{Tulin:2013teo}. 
	From Bullet clusters, the bound $\sigma(\chi \chi \to \chi \chi)/m_\chi<1.25$ cm$^2$/gr has been derived \cite{Clowe:2006eq,Randall:2007ph}.
	For non-chiral interaction with $b=0$, we can write  $\sigma(\chi \chi \to \chi \chi)\sim g_\chi^4/(4\pi m_\chi^2 v_\chi^4)$ \cite{Ackerman:mha} so   the bound $\sigma(\chi \chi \to \chi \chi)/m_\chi <1 ~{\rm cm}^2/{\rm gr}$ for $\chi$ with velocity of $v_\chi\sim 10^{-3}$ in the galaxy  implies $g_\chi<0.05$  which is  readily satisfied in the perturbative range. 
	However, for $b\ne 0$ the cross section $\sigma (\chi\chi\to \chi \chi)\sim b^4 g_\chi^4 m_\chi^2/(4\pi m_{Z'}^4)$ is greatly enhanced by  $(m_\chi^4/m_{Z'}^4)$. The enhancement comes from the $q_\mu q_\nu/m_{Z'}^2$ part of the $t$- and $u$-channel propagator and the fact while $q_\mu \bar{\chi}_2 (q+p)\gamma^\mu \chi_1(p)=0$, the axial part does not vanish: $q_\mu \bar{\chi}_2 (q+p)\gamma^\mu \gamma^5\chi_1(p)=m_\chi \bar{\chi}_2 (q+p) \gamma^5\chi_1(p)$. To avoid too strong self-interaction, we set $b=0$. That is we take the coupling to dark matter to be non-chiral.

	In principle, $\chi \chi \to \chi \chi Z'$ can lead to dissipation. To prevent this, we should require $\sigma (\chi \chi \to \chi \chi Z')  v_\chi n_\chi t_0<1$ where $t_0 \sim 10^{17}{\rm sec}$, $n_\chi={\rm cm}^{-3}(\rho_\chi/5 ~{\rm GeV}{\rm cm}^{-3})(5~{\rm GeV}/m_\chi)$ and $\sigma( \chi \chi \to \chi \chi Z') v_\chi \sim [g_\chi^4/(4\pi m_{Z'}^2)]\times [10 g_\chi^2/16\pi^2]$ where the factor of 10 takes care of the IR logarithmic enhancement.   The $m_{Z'}^2$ dependence comes from the longitudinal component of $Z'$. The dissipation constraint then leads to $g_\chi<0.008$  which is  again satisfied within the perturbative range.  }


Finally let us discuss the possibility that at temperatures higher than the electroweak symmetry breaking new processes involving $\chi$
 lead to thermalization of $\nu_s$ and $Z'$.
The densities of $\nu_s$ and $Z'$ will be diluted because of the entropy dump  in plasma
so their contribution to  $N_{eff}$ will be negligible, avoiding the bounds  from BBN and CMB.
At temperatures MeV$<T<$ 100~MeV where $\Delta m_{21}^2/T \gg H$, the active neutrinos are mainly in form of incoherent (effective) mass
 eigenstate, $\nu_i$ which have a coupling of form $\bar{\nu}_i\gamma^\mu \nu_s Z'_\mu$ given by $g_s U_{si} [\Delta m_{41}^2 /(2E_\nu V_s)]$.
The matter potential due to ADM is
\be
V_{s} = \frac{g_{s}g_{\chi}}{m_{Z'}^{2}}\eta_{\chi} n_\gamma
\ee
where $\eta_{\chi}$ is the DM asymmetry parameter; $\eta_\chi \equiv (n_{\chi}- n_{\bar{\chi}})/n_{\gamma}$. Since $\Omega_{\chi} \simeq 5 \Omega_{B}$, one typically requires $m_{\chi} \eta_{\chi} = 5 m_{p} \eta_{B}$, suggesting $m_{\chi} \simeq 5 $ GeV when $\eta_{B} = \eta_{\chi}$.
 Let us check if the scattering of $\nu_a$ off relic $\nu_s$ can populate $\nu_s$ at $T>1$ MeV via $\nu_a+\nu_s \to \nu_s+\nu_s$.
 The rate of scattering of each relic $\nu_s$ can be written as $\Gamma =\sum n_\nu \sigma_s$ where for $T\gg m_{Z'}$, the process
 has cross section of $\sigma_s \sim (g_s^4 U_{si}^2/4\pi T^2) (\Delta m_{41}^2/2T V_s)^2$.
 In order for $\Gamma < H \sim T^{2}/M_{Pl}^*$ at $T \sim$ MeV, we find that we need
 \be \label{chi-s}3\times 10^{-5}\left( \frac{\Delta m_{41}^2}{e{\rm V}^2}\right)  \left( \frac{m_{Z'}}{10~e{\rm V}} \right)^2 g_s<g_{\chi},\ee
 which for values of $g_s$ and $g_\chi$ giving rise to $\epsilon_\chi >10^9$ can be readily satisfied.

In summary, combining the conditions of decay of $\nu_4$ before recombination and free streaming of light neutrinos at that epoch implies
$$ 3\times 10^{-4} \left(\frac{0.1}{U_{si}}\right)^{3/2} \lesssim g_s \lesssim 5 \times 10^{-4}  \left(\frac{0.1}{U_{si}}\right)^{2}.$$
Moreover, the suppression of $\nu_s$ (and therefore of $Z'$ production) before neutrino decoupling implies $\epsilon_\chi \gg 10^9$.

{\it Bounds from neutrino oscillation experiments:} As mentioned in the introduction, the MiniBooNE collaboration has suggested resonance matter effect with $E_{res}\sim 300$ MeV as a solution
to the  MiniBooNE anomaly. This value of resonance for energies and densities  of interest to MiniBooNE corresponds $\epsilon \sim 10^4$.\underline{}
Such large $\epsilon$ within our model requires $g_s \sim 10^{-3}$ and $g_B\sim 10^{-13}-10^{-12}$ at $m_{Z'}\sim 10$ eV which are allowed.
However, with $\epsilon\sim 10^4$, the effective mass splitting between active neutrinos for $E_\nu>1$ GeV obtains a correction given by  $\Delta m_{41}^2(1-|U_{s 4}|^2)$ \cite{Karagiorgi:2012kw}
 which is much larger than $\Delta m_{31}^2$. Thus, the pattern of oscillation in long baseline and atmospheric neutrino data will be dramatically changed, constraining $\epsilon$ to values smaller than $O(10)$ \cite{Kopp:2014fha}.
 
 {\it Impact on supernova evolution:} The coupling of $Z^\prime$ particles to standard model particles is too small  to allow an efficient
production of $Z^\prime$ in a supernova. If $\nu_s$ particles exist inside the supernova, they can produce $Z^\prime$ particles. However, inside the supernova the matter effects suppress the effective mixing between active and sterile neutrinos.
Moreover, since the rate of scattering of active neutrinos off nuclei via electroweak interaction is much larger that oscillation time, active neutrinos stay in coherent active form with no coupling to $Z'$ so processes such a $\nu_a \bar{\nu}_a \to Z'Z'$ or $\nu_s \bar{\nu}_s$ cannot take place\footnote{There is a subtlety here. Although the mass eigenstate $\nu_i$ can have coupling to $Z'$ and $\nu_s$ given by $g_s U_{si}$, there is no coupling of form $\bar{\nu}_a \gamma^\mu \nu_s Z'_\mu$.}.
Thus, there is no possibility of $\nu_s$ production in the core. When the active neutrinos or antineutrinos stream out of the core and reach densities of $O[10^4~{\rm gr/ cm}^3 (10/\epsilon)]$, depending on the sign of $V_s$, they may undergo resonant conversion to sterile neutrinos which will leave its imprint in the spectrum of neutrinos observed on the earth. Studying the whole effect is beyond the scope of the present paper.

{\it Conclusion:} Achieving $\epsilon\sim 10$ and $\epsilon_\chi \gg 10^9$ is possible with the values of coupling satisfying the present bounds.
	For example,
	taking $g_\chi \sim 3 \times 10^{-5}$ (to remain in the perturbative range), $g_s \sim 3 \times 10^{-4}$ $g_B\sim 2 \times 10^{-16}$ and $m_{Z^\prime}\sim 10$ eV, we obtain
	\be \epsilon =\frac{g_sg_B}{6\sqrt{2}m_{Z^\prime}^2 G_F}=6 \ \ \ \ {\rm and} \ \ \ \ \epsilon_\chi =\frac{g_sg_\chi}{2\sqrt{2}m_{Z^\prime}^2 G_F}= 10^{12}.  \label{eps} \ee
	With such values of $\epsilon_\chi$, the active sterile mixing is suppressed down to 300~keV$( 10^{12}/\epsilon_\chi)^{1/4}$.  Below $T\sim 300$~keV (well below the decoupling of neutrinos) active neutrinos can oscillate to $\nu_s$.
	Then, $Z'$ and  $\nu_s$ come to equilibrium with active neutrinos again via $\sigma (\nu \bar{\nu}\to \nu_s \bar{\nu}_s)\sim \sigma (\nu \bar{\nu}\to Z' Z') \sim g_s^4U_{si}^4 T^2/(4\pi)$ and subsequent $Z'\to \nu_s \bar{\nu}_s$. As explained above, at $T<m_4\sim 1$ eV
	(when structure formation starts) neutrinos resume free streaming.

\section{The Model\label{model}}

In this section, we discuss how to build a $U_X(1)$ gauge model with gauge boson of mass $\sim 10$~eV that can give rise to sizeable non-standard interactions
of sterile neutrinos with matter fields. We shall call this model the $3+1+U(1)$ model.
Let us take the gauge coupling equal to $g_B$ and assign the following $U_X(1)$ charges to the standard model fermions:

\be B+a_e L_e+a_\mu L_\mu +a_\tau L_\tau \ . \label{B+LLL}\ee
Notice that we have assigned equal $U_X(1)$ charges to quarks of all three generations. Had we assigned unequal charges to quarks of different flavors,
$Z^\prime$ would have obtained couplings of form $Z^\prime_\mu \bar{q}_i \gamma^\mu q_j$ where $q_i$ and $q_j$ are quarks of different masses.
 Such a coupling could lead to $q_i \to q_j Z^\prime$ enhanced by $(m_{q_i}/m_{Z^\prime})^2$. As long as $a_e+a_\mu+a_\tau=-3$, the chiral anomalies cancel out. For $a_e+a_\mu+a_\tau \ne -3$, new chiral fermions charged under $U_X(1)$ should be added to the standard model to cancel the anomalies.

The effective coupling  between quarks and active neutrinos will be then  of form $$2\sqrt{2}\epsilon_{\alpha \alpha}G_F(\bar{\nu}_\alpha \gamma^\mu P_L \nu_\alpha)(\bar{q} \gamma_\mu q) $$ where $\epsilon_{\alpha \alpha} =a_\alpha  (g_B^2/m_{Z^\prime}^2)(1/6\sqrt{2}G_F)$.
In the parameter range of our interest, the NSI couplings of active neutrinos can be estimated as $\epsilon_{\alpha \alpha} \sim 10^{-12}a_\alpha (g_B/10^{-16})^2(10~e{\rm V}/m_{Z^\prime})^2$.  The non-standard interactions of active neutrinos will be therefore too small to be observable. Remember that in our scenario,  active neutrinos are not  supposed to have sizeable NSI.

 The sterile neutrino  is taken to be  an electroweak singlet  fermion with   $U_X(1)$ charge equal to $a_s=g_s/g_B$.
 However, mixing between $\nu_s$ and $\nu_\alpha$ breaks both $SU(2)\times U(1)$ and $U_X(1)$ which can be achieved by the mechanism described below \cite{Farzan:2016wym}:
Let us introduce a new singlet
Weyl fermion $N_R$ neutral under both the SM and  $U_X(1)$ and a new scalar
$\phi$ which is charged only under  $U_X(1)$ with a charge equal to that of $\nu_s$. We can then write the following Yukawa couplings
$$ \lambda_{aN}\bar{N}_R H^T\epsilon L + \lambda_{sN} \phi \bar{\nu}_s N_R+{\rm H.c.} $$ where $H$ is the SM Higgs. VEVs of $H$ and $\phi$ respectively induce masses of form $m_{aN}=\lambda_{aN} \langle H\rangle$ and $m_{\nu_s}=\lambda_{sN}\langle \phi\rangle$. 
Taking $m_{aN}/m_{\nu_s}\sim U_{a4}\sim 0.1$ and $m_{\nu_s}\sim 1$ eV, we obtain $\lambda_{a N}\sim 10^{-12}$ so the branching ratio of invisible decay mode $H\to \nu N$ is quite suppressed. Moreover this interaction at early universe cannot bring $N_R$ to thermal equilibrium $\lambda_{aN}^2T \ll H=T^2/M_{Pl}^*$ for $T\gtrsim 200$ GeV.
The mass term can be written as \begin{eqnarray} [\nu_a^T \ \nu_s^T \ N^{cT}] (i\sigma_2) \left[ \begin{matrix} m_a & 0 & m_{aN} \cr
0 & 0 & m_{\nu_s} \cr m_{aN} & m_{\nu_s} &0 \end{matrix} \right] \left[ \begin{matrix} \nu_a \cr \nu_s \cr N^c \end{matrix} \right],\end{eqnarray} where  $m_a$ is the mass of active neutrinos, which might be produced by any of the mechanims introduced in the literature. Taking $m_a^2\ll m_{aN}^2\ll m_{\nu_s}^2$, the mass matrix can be diagonalized via \begin{eqnarray} O=\left[\begin{matrix} 1 & -\frac{m_{aN}}{m_{\nu_s}} &0 \cr \frac{m_{aN}}{\sqrt{2} m_{\nu_s}} & \frac{1}{\sqrt{2}}&-\frac{1}{\sqrt{2}}   \cr \frac{m_{aN}}{\sqrt{2} m_{\nu_s}} & \frac{1}{\sqrt{2}} &\frac{1}{\sqrt{2}} \end{matrix} \right],\end{eqnarray} with the mass eigenvalues equal to $\{ m_a ,-m_{\nu_s},m_{\nu_s}\}$.  Notice that up to  corrections of $O\left((m_{aN}/m_{\nu_s})^3\right)$, $m_a$ will not receive corrections.
In other words, the contribution from ``seesaw-like" mechanism to active neutrino mass is $m_{\nu_s}(m_{aN}/m_{\nu_s})^3$ which is much smaller than $\sqrt{\Delta m_{atm}^2}$ so $m_a$
should originate from another mechanism.
Moreover, for the first approximation $\nu_a$ does not mix with
$N_R$. The interesting point is that even after turning on $V_s$, $\nu_a$ does not mix with $N^c$. This can be understood as follows. The Weyl fermions $\nu_s$ and $N_R$ together form a Dirac fermion, $\psi$ where $\psi_L=\nu_s$ and $\psi_R=N_R$. Regardless of whether we turn on matter effects, only $\psi_L$ will be involved in oscillation. This is a well-known result which comes from the fact the oscillation is given by $m_\nu^\dagger.m_\nu$ rather than by $m_\nu$. The VEV of $\phi$ gives a contribution to $m_{Z'}^2$ given by $g_s^2\langle \phi\rangle^2$. Taking $m_{\nu_s}=\lambda_{sN}\langle \phi\rangle \sim 1$ eV, the condition $\langle \phi \rangle \leq m_{Z'}/g_s=33~ ke{\rm V}(m_{Z'}/10~ e{\rm V})(3\times 10^{-4}/g_s)$ implies $\lambda_{sN} \geq 3 \times 10^{-5}$. Taking the quartic self coupling of $\phi$ to be
$\lambda_\phi$,  it is natural that the mass of $\phi$ to be $m_\phi \sim \sqrt{\lambda_\phi} \langle \phi \rangle <33~keV$. Taking $\lambda_\phi \sim 0.1-1$, $\phi$ particles can be produced in the early universe at temperatures around 100 keV when $\nu_s$ come into equilibrium with active neutrinos. {At these temperatures the $\phi$ mass obtains a correction of $\lambda_{sN} T \sim (m_{\nu_s}/\langle \phi \rangle)T$ which is smaller than $\langle \phi \rangle$ and $m_\phi$ for $\langle \phi \rangle >300$ eV and therefore negligible.} When the temperature drops below $m_\phi$, $\phi$
decays to sterile neutrinos, $N$ and $\nu_s$ which in turn decay into lighter neutrino states.  As mentioned above, $N$ and $\nu_s$ form a Dirac fermion whose left-handed component has a $U_X(1)$ charge. This induces a $U_X(1)-U_X(1)-U_X(1)$ chiral anomaly. As discussed in the previous section, the coupling of the dark matter field, $\chi$, has to be non-chiral so it cannot help with anomaly cancellation. To cancel anomaly, we may add another chiral field, $\nu_R$ with 
$U_X(1)$ charge equal to those of $\nu_s$ and $\phi$. If the mass of $\nu_R$ comes from a VEV of a scalar charged under $U_X(1)$, it cannot be heavier than $\sim 30$ keV.  $\nu_R$ can have a small coupling of form $\lambda_{RN} \phi^\dagger \nu_R^T cN_R$ with $\lambda_{RN}\ll \lambda_{sN}$, providing it with a decay mode  so in case it is produced in the early universe, it can decay away immediately.

As mentioned before, with $a_e+a_\mu+a_\tau=-3$, chiral anomalies involving the SM fermions cancel without a need for extra fermionic degrees of freedom which are chiral under the gauge group.
The $B-L$ symmetry
({\it i.e.,} taking $a_e=a_\mu=a_\tau=-1$ ) is compatible  with mixing in the lepton sector. That is obtaining a mixing between different neutrino flavors does not require an extra scalar whose VEV breaks $U_X(1)$. Of course, anomaly cancellation requires adding the right-handed neutrinos. These right-handed neutrinos can help to give Dirac mass to neutrinos. If no Majorana mass (which breaks $U_X(1)$) is induced, these right-handed neutrinos will be degenerate with light active neutrinos and can kinematically be produced in the early universe. Taking $g_B \lesssim 10^{-13}$, they can never come to thermal equilibrium: $\Gamma(\nu_L\bar{\nu}_L \to \nu_R\bar{\nu}_R) \sim (a_\alpha g_B)^4T/(4\pi)|_{T\sim {\rm MeV}} \ll H=T^2/M_{Pl}^*|_{T\sim {\rm MeV}}$.
It is also possible to implement seesaw mechanism by introducing an extra scalar whose $U_X(1)$ charge is equal to $-2$ times $U_X(1)$ charge of leptons $S^\prime\nu_R^Tc\nu_R$. The upper bound on $\langle S^\prime\rangle$ from the mass of $Z^\prime$ is $100~{\rm TeV} (m_{Z^\prime}/10~{\rm eV})(10^{-13}/g_B)$.

Since in the electrically neutral mediums such as the Earth, the number densities of the electrons and the proton are equal, for the case
of gauging the $B-L$ symmetry, the contributions from the electron and proton  to $V_s$ cancel out and $V_s$ turns out to be proportional to
the neutron number density:
\be \label{Vs2} V_s=3(2{\sqrt{2}})G_FN_n \epsilon ,\ee
where the factor of 3 reflects the fact that there are 3 quarks in a neutron\footnote{Note that $\eps$ defined in this text is the ``Lagrangian" level NSI, as opposed to the ``Hamiltonian" level NSI often also used in the literature. These differ by a factor of six: three from the quarks and two since we do not include a $P_L$ operator in front of matter field,
	$f$,
	in Eq.~\ref{s-bar} and only consider vector NSI.}. This factor of 3 compensates the factor of $1/3$ in Eq.~(\ref{eps}).

{Lastly note that if the VEV is zero until very late times (i.e. $T < $ MeV), then the cosmological bounds are also significantly weakened. This is because the presence of a nonzero VEV is required in order for the active and sterile neutrinos to mass mix, and the only mechanism for thermalization for the steriles is via mixing. These scenarios have been explored in~\cite{Patwardhan:2014iha,Vecchi:2016lty}. }

{\it Summary:}
We presented a simple anomaly free Abelian gauge $U_X(1)$ model based on gauging $B-L$. The
$\nu_s$, having arbitrary charge under this new $U_X(1)$ is the left-handed component of a Dirac fermion. The right-handed component is neutral under gauge symmetry and can have Yukawa coupling with SM neutrinos. The mass of the Dirac fermion comes from the VEV of light (keVish) scalar, $\phi$,
charged under the new   $U_X(1)$. The parameter range of interest comes out naturally without a need for fine-tuning. In Table~\ref{tab:symbols}
we present a fiducial summary of parameters needed in the model consistent with all bounds.

\begin{table}
\begin{tabularx}{\textwidth}{ |X|X|X| }
  \hline
  {\bf Quantity }& {\bf  Symbol }&{\bf  Fiducial Value}   \\
  \hline 
  $Z'$ mass  & $m_{Z'}$ & $10$ eV   \\
\hline
  Sterile-$Z'$ coupling  & $g_{s}$ & $3 \times10^{-4}$  \\
\hline
  Baryon-$Z'$ coupling  & $g_{B}$ & $2 \times10^{-16}$  \\
\hline
  $\phi  $ VEV  & $\langle \phi \rangle $ & $30$ keV  \\

  \hline
\end{tabularx}
\\
\caption{Summary of notation and key fiducial values in the model.   }
\label{tab:symbols}
\end{table}

\section{The Oscillation Picture}
\label{osc}
Despite the 6~$\sigma$ evidence for a $\sim1$ eV sterile neutrino from LSND and MiniBooNE along with other weaker hints, there is compelling evidence from IceCube, MINOS, MINOS+ against the $3+1$ scenario with a  $\sim1$ eV sterile neutrino.
In this section we address whether  the addition of a new interaction in the sterile sector modifies these constraints.

In the $3+1$ scenario,
the unitary mixing matrix $U$ is the PMNS matrix \cite{Pontecorvo:1967fh,Maki:1962mu}  extended to include a fourth generation via $U=R_{34}(\theta_{34})R_{24}(\theta_{24})R_{14}(\theta_{14})U_{\rm PMNS}$.
We choose this definition to make the $U_{e4}=\sin\theta_{14}$ and $U_{\mu4}=\cos\theta_{14}\sin\theta_{24}$ terms (which are relevant for LSND and MiniBooNE) as simple as possible.
We have taken all new CP violating phases to be zero for simplicity.

Section \ref{ATMice} describes the predictions of the $3+1$ and $3+1+U(1)$ scenarios for high energy atmospheric neutrinos at IceCube. Section \ref{sec:oscillation results} summarizes IceCube results for the $3+1+U(1)$ scenario. Section \ref{MINOSbound} summarizes the bounds from MINOS and MINOS+ on the  $3+1+U(1)$ scenario.
Section \ref{ComPic} discusses the bounds from IceCube and MINOS/+ allowing all the mixing angles including $\theta_{34}$ to be nonzero.

\subsection{The Atmospheric Neutrino Constraint from IceCube\label{ATMice}}
\begin{figure}
\includegraphics[width=0.49\textwidth]{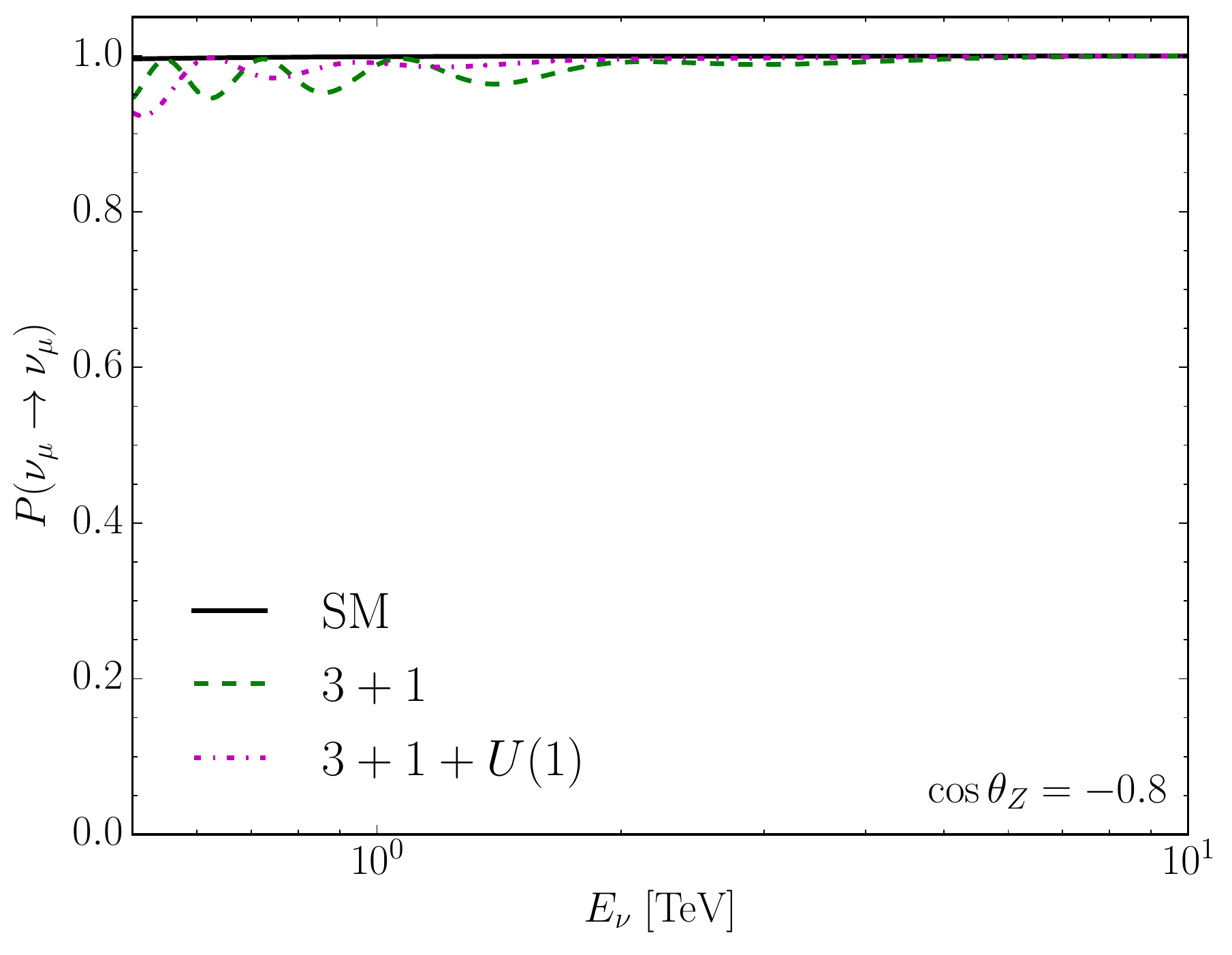}
\includegraphics[width=0.49\textwidth]{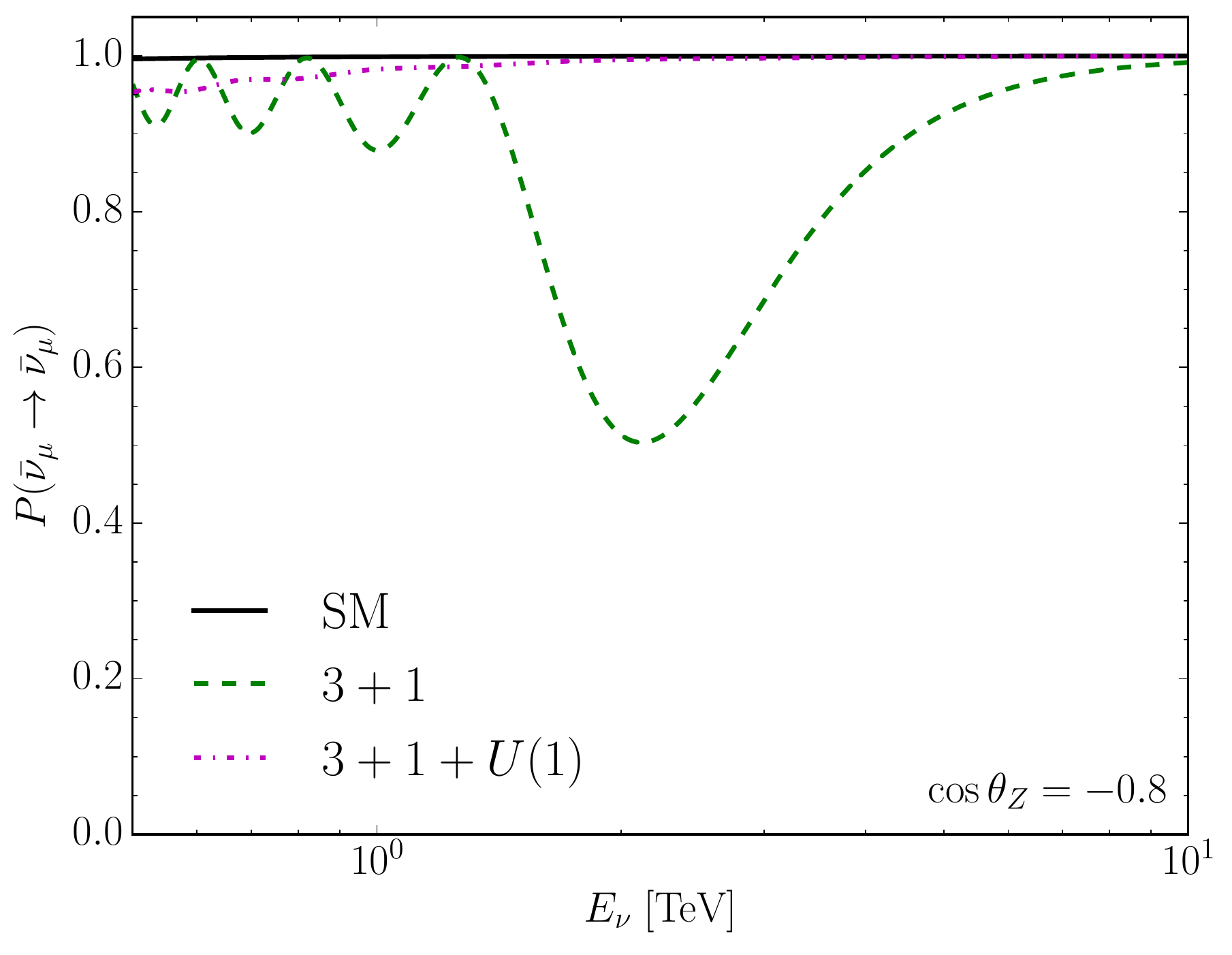}
\caption{The $\nu_\mu\to\nu_\mu$ survival probabilities at $\cos\theta_Z=-0.8$ for three different cases.
For the SM the standard oscillation parameters are taken from \cite{Esteban:2016qun,nu-fit:v3.2}.
For the other cases we take our benchmark sterile parameters: $\theta_{14}$ from reactor data and $\theta_{24}$ from a global fit including LSND and MiniBooNE.
The NSI case is evaluated at our best fit point from IceCube (see section \ref{sec:oscillation results} below) $\epsilon=-1.3$.}
\label{fig:probabilities}
\end{figure}

IceCube has provided one of the strongest constraints on $\sim1$ eV sterile neutrinos by measuring  atmospheric $\nu_\mu$ disappearance probabilities at energies $\gtrsim1$ TeV  \cite{Aartsen:2017bap}. Beyond these energies, the oscillation length becomes larger than the Earth diameter so neutrinos are not expected to oscillate in the Earth.
Thus, at $E\gtrsim1$ TeV without any new physics we have $P(\nu_\mu\to\nu_\mu)\approx1$ as shown in black in fig.~\ref{fig:probabilities}.
The existence of both a $\sim1$ eV sterile neutrino and NSI's will alter this as was shown in \cite{Liao:2016reh}.

We calculate the effect of a sterile neutrino with a new interaction as it would affect IceCube's measurement.
For the matter density profile we use the Preliminary Reference Earth Model \cite{Dziewonski:1981xy} and calculate the oscillation probabilities through the Earth numerically.
Phenomenologically, our model contains five new parameters beyond the standard oscillation parameters: $\Delta m^2_{41}$, $\theta_{14}$, $\theta_{24}$, $\theta_{34}$, and $\eps$.
For the standard oscillation parameters we use the result of the global fit and assume the normal mass ordering (which is preferred at $\sim3$ $\sigma$) from \cite{Esteban:2016qun,nu-fit:v3.2}.

LSND and MiniBooNE interpret their results in a two flavor picture where the probability is given by
\begin{equation}
P(\nu_\mu\to\nu_e)=1-\sin^22\theta_{\mu e}\sin^2\left(\frac{\Delta m^2L}{4E}\right)\,.
\end{equation}
This mixing angle is related to the angles given above by $\sin^22\theta_{\mu e}=\sin^2\theta_{24}\sin^22\theta_{14}$.

We fix $\sin^2\theta_{14}=0.095$ which is the best fit point of the global fit to reactor $\nu_e$ disappearance without input from theoretical reactor fluxes \cite{Dentler:2018sju}.
While a recent analysis from Daya Bay suggests that the Reactor Antineutrino Anomaly (RAA) may be due (in part or in full) to nuclear effects \cite{An:2017osx}, a more recent article indicates that more analysis is required \cite{Hayes:2017res}.
Regardless, even in the event that the RAA is due to nuclear effects we take $0.095$ as a reasonable upper limit on $\sin^2\theta_{14}$.
We want $\theta_{14}$ large to minimize the amount of sterile mixing in the $\nu_\mu$ sector where the IceCube (and MINOS and MINOS+ below) constraints are strong.

For the remaining two sterile parameters, we define our benchmark values based on a global fit to the $\nu_\mu\to\nu_e$ data including LSND's decay-in-flight data, which gives $\sin^22\theta_{24}=0.0664$ and $\Delta m^2_{41}=0.559$ eV$^2$ \cite{Dentler:2018sju}.
While this does not include the latest MiniBooNE results, the new results do not change the region of interest much, only the significance.

Figure \ref{fig:probabilities} shows the oscillation probabilities for the SM, our benchmark sterile neutrino only model ({\it i.e.,} $3+1$), and our benchmark sterile neutrino model with NSI ({\it i.e.,} $3+1+U(1)$) fixed to what will be our best fit result from IceCube.
The large resonant oscillation probability for anti-neutrinos in the $3+1$ model is the result of the MSW resonance \cite{Wolfenstein:1977ue,Mikheev:1986gs} in the Earth, and is the signal that IceCube is particularly sensitive to.
The addition of large NSI significantly suppresses the resonant conversion and mostly returns the probability to that in the standard $3\nu$ scheme.

\begin{figure}
\centering
\includegraphics[width=0.49\textwidth]{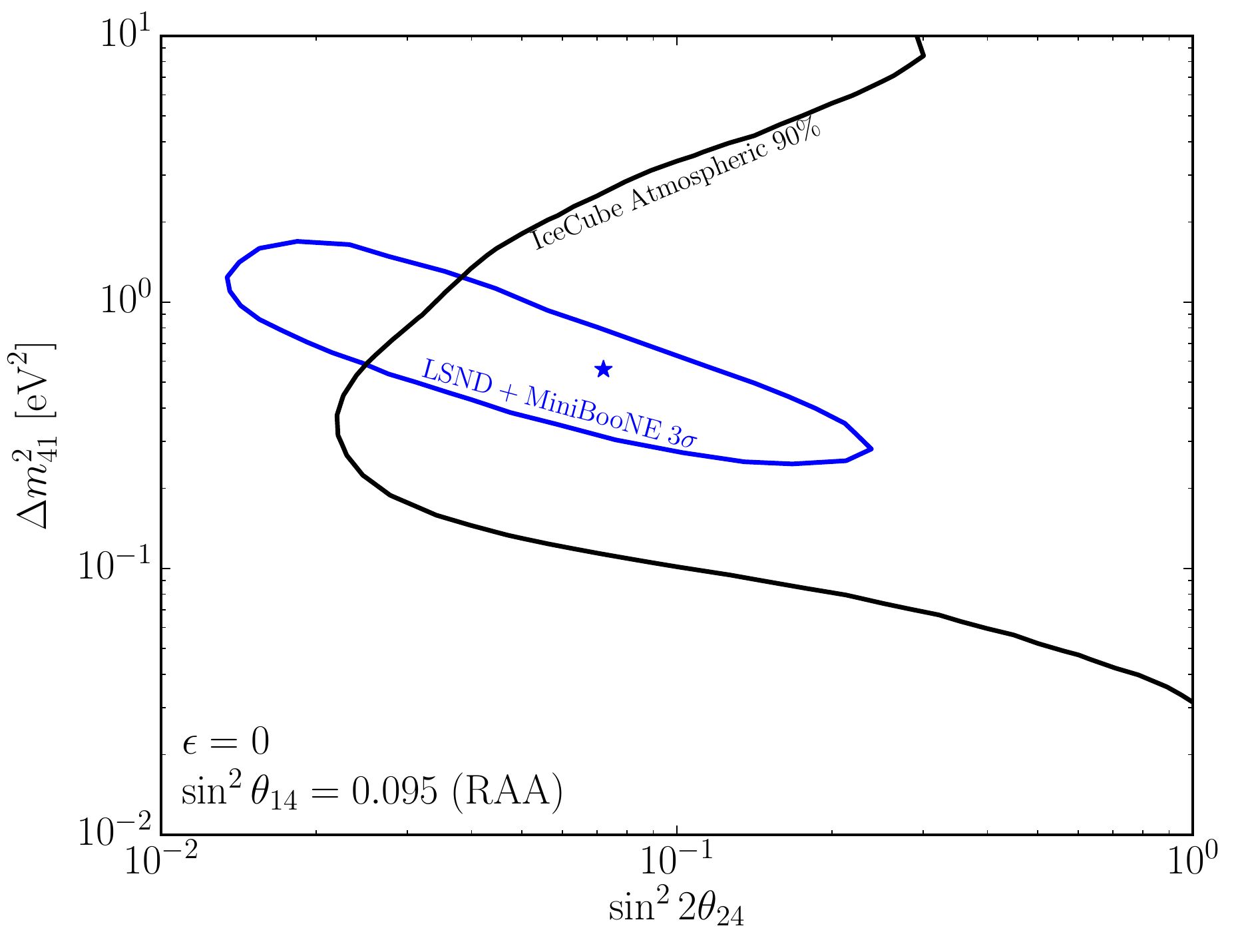}
\includegraphics[width=0.49\textwidth]{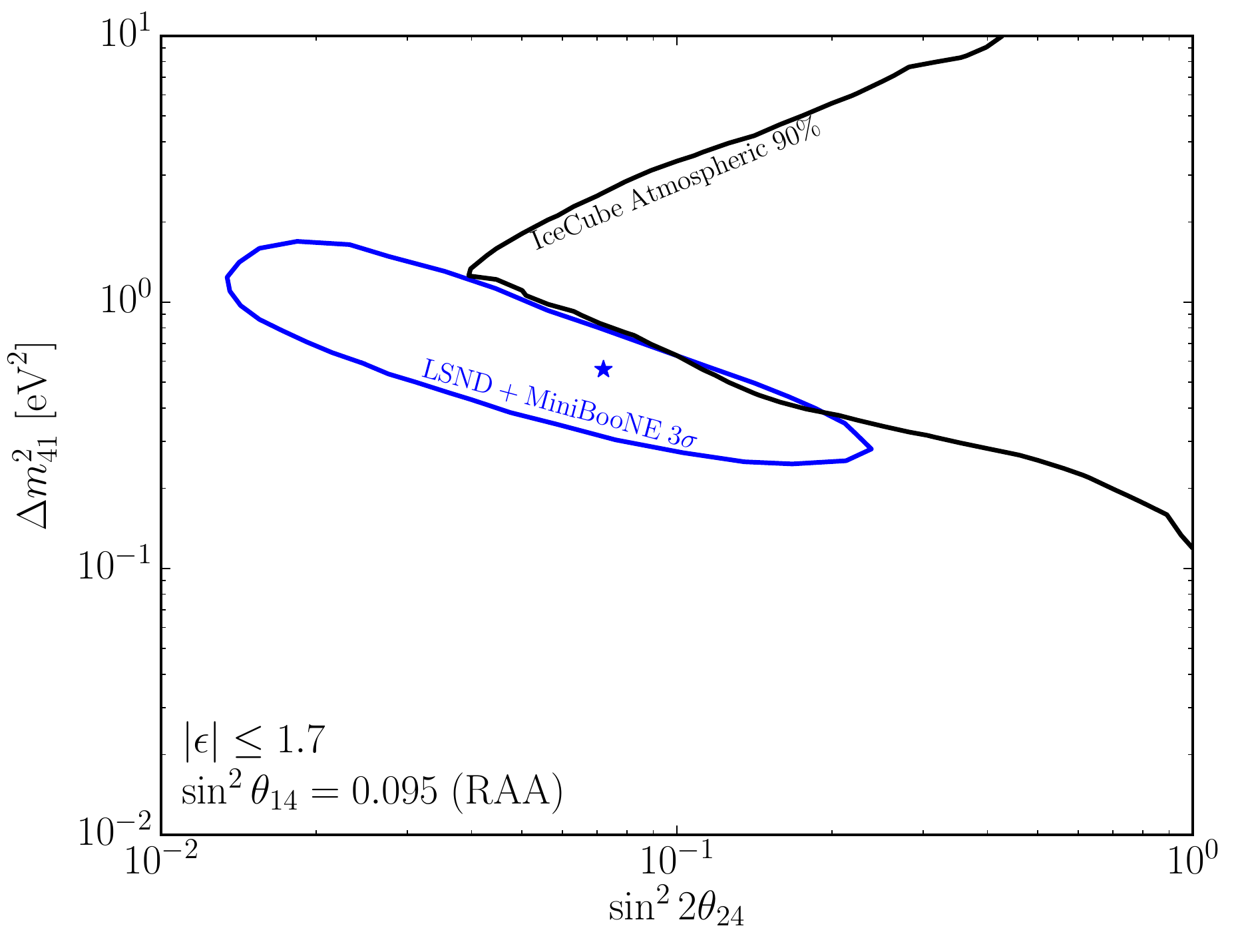}
\caption{The 90\% exclusion region from IceCube atmospheric neutrino data with $500~{\rm GeV}<E_\nu<10$ TeV for 2 degrees of freedom is shown as a function of $\Delta m^2_{41}$ and $\sin^22\theta_{24}$ where $\theta_{14}$ is fixed to the best Reactor Antineutrino Anomaly value.
Also shown is the 3 $\sigma$ allowed region from the global fit to $\nu_\mu\to\nu_e$ appearance data including LSND's decay-in-flight data; the best fit point is denoted with a star \cite{Dentler:2018sju}.
On the left is the sterile only model ({\it i.e.,} $3+1$), and on the right an additional minimization is performed over $|\epsilon|\le1.7$ for the $3+1+U(1)$ scenario.}
\label{fig:scan2D}
\end{figure}

We use the publicly available two year upward-going muon neutrino flux from IceCube with $\sim35,000$ events \cite{Aartsen:2015rwa,IC:HE_NuMu_diffuse} with $501$ GeV $<E_{\nu,{\rm p}}<10$ TeV and $\cos\theta_Z<0$ binned into 13 energy bins and 10 angular bins where $E_{\nu,{\rm p}}$ is the energy proxy used by IceCube.
The atmospheric flux is provided by \cite{Honda:2004yz}.
It is extended to the TeV range as described in \cite{Aartsen:2013eka} using the cosmic ray flux provided in \cite{Gaisser:2011cc} to account for the knee, and there is an estimated uncertainty of $\sim25\%$ at $E_\nu\sim1$ TeV.
IceCube's effective area as a function of energy, angle, and flavor along with the Digital Optical Module (DOM) efficiency is provided by IceCube \cite{IC:HE_NuMu_diffuse} and then used to calculate the expected number of events in each bin:
\begin{equation}
N(E_{\nu,{\rm p}},\cos\theta_Z)=\int dE_{\nu,{\rm p}}\int d\cos\theta_Z\int dE_\nu A(E_{\nu,{\rm p}},E_\nu,\cos\theta_Z)\Phi(E_\nu,\cos\theta_Z)P(E_\nu,\cos\theta_Z)\,,
\end{equation}
where $E_\nu$ is the true neutrino energy, the $E_{\nu,{\rm p}}$ and $\cos\theta_Z$ integrals are taken over the bin size, $A$ is IceCube's effective area which includes  DOM efficiency, $\Phi$ is the atmospheric flux, and $P$ is the $\nu_\mu$ disappearance probability.
We then construct a $\chi^2$ test statistic,
\begin{equation}
\chi^2=\min_x\left\{2\sum_i\left[(1+x)N_{i,{\rm th}}-N_{i,{\rm IC}}+N_{i,{\rm IC}}\log\frac{N_{i,{\rm IC}}}{(1+x)N_{i,{\rm th}}}\right]+\left(\frac x{\sigma_x}\right)^2\right\}\,,
\end{equation}
where the sum is over the $E_{\nu,{\rm p}}$ and $\cos\theta_Z$ bins, and we have accounted for the atmospheric flux normalization uncertainty $\sigma_x=0.25$ with a pull term \cite{Fogli:2002pt}.

\subsection{IceCube Results}
\label{sec:oscillation results}
We perform a scan over the two sterile parameters $\theta_{24}$ and $\Delta m^2_{41}$ for the case with no NSI ({\it i.e.,} the 3+1 scheme) and for the $3+1+U(1)$ scenario where we minimize over $|\eps|\le1.7$. The results are shown  in fig.~\ref{fig:scan2D}.
While the sterile neutrino picture is disfavored by the IceCube data, we find that the addition of NSI with $|\eps|\le1.7$ describes the data with 500~GeV$<E_\nu<10~{\rm TeV}$ well.

\begin{figure}
\centering
\includegraphics[width=4in]{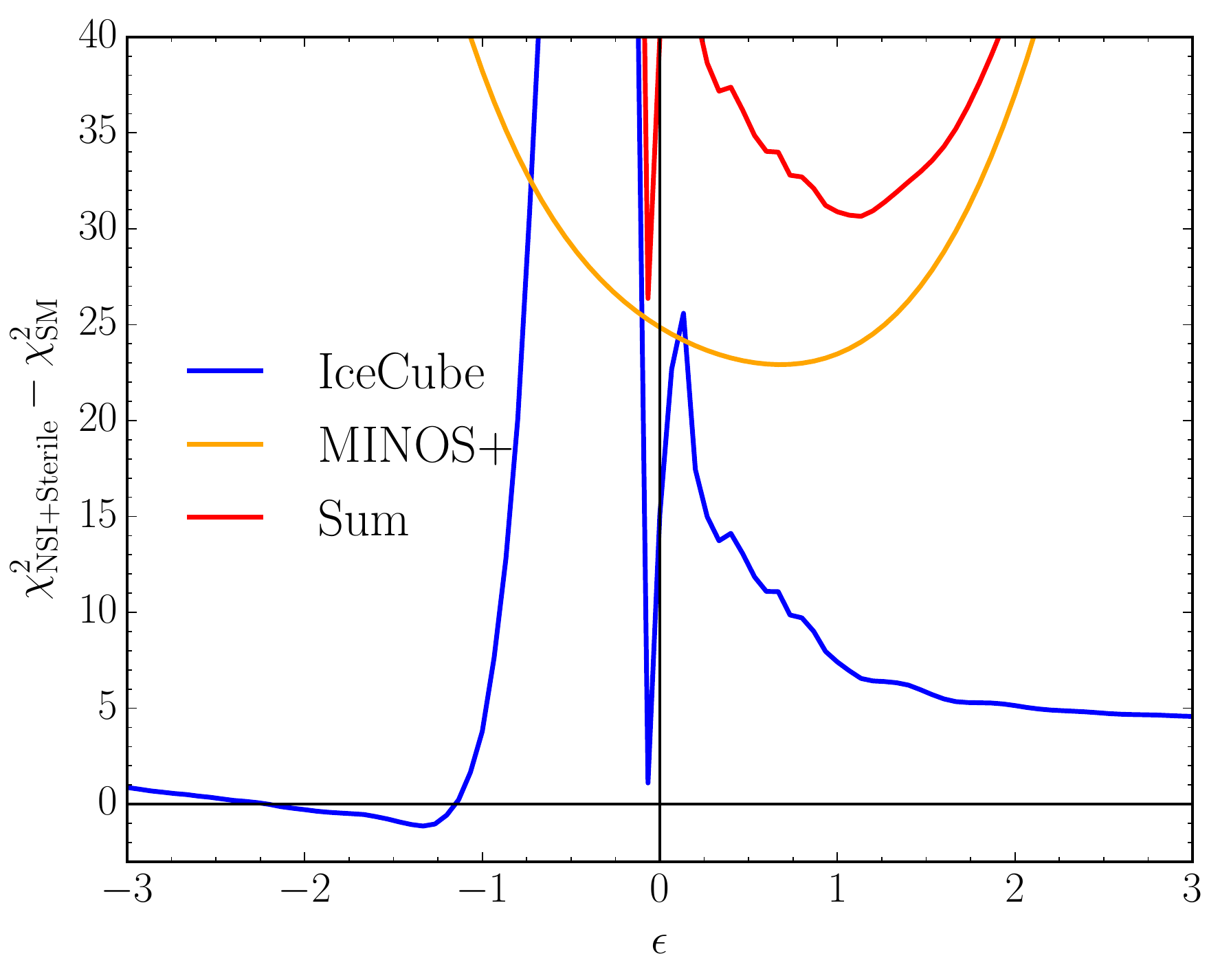}
\caption{The $\Delta\chi^2$ as a function of $\epsilon$ where the sterile parameters have been set to their best fit values from \cite{Dentler:2018sju} with $\theta_{34}=0$.
The 3+1 model at $\epsilon=0$ is disfavored at $\chi^2=15.1$ for IceCube atmospheric neutrino data with $500~{\rm GeV}<E_\nu<10$ TeV and $24.9$ for MINOS and MINOS+.
There is a small region where sterile neutrinos and NSI are actually slightly favored over the SM by IceCube with a best fit value of $\epsilon=-1.3^{+0.2}_{-0.8}$.
For MINOS and MINOS+ the picture is never better than the SM and the improvement over the 3+1  picture is only marginal.
The best fit point for both experiments is at $\eps=-0.07$ at $\Delta\chi^2=26$ which is preferred over the best fit $3+1$ (at $\Delta\chi^2=40$) by an improvement of $\Delta\chi^2=14$.}
\label{fig:GF chisq}
\end{figure}

Next, we fix the sterile parameters to their best fit values from \cite{Dentler:2018sju} and vary only the NSI parameter $\eps$ in fig.~\ref{fig:GF chisq}.
The $3+1$ hypothesis is disfavored compared to the standard $3\nu$ scheme at $\Delta\chi^2=15.1$.
The addition of NSI in the sterile sector not only relaxes this tension, but provides a slightly better ($\sim1$ $\sigma$) fit to the data with 500~GeV$<E_\nu<10~{\rm TeV}$ for $\eps=-1.3^{+0.2}_{-0.8}$ where the error bars are the 1 $\sigma$ uncertainty.

Finally, we plot the sum of both experiments in red.
We find that the best fit point for both experiments for $3+1+U(1)$ is at $\eps=-0.07$ with $\Delta\chi^2=26$ compared to the SM.
The $3+1+U(1)$ is preferred over the best fit $3+1$ point (at $\Delta\chi^2=40$) by an improvement of $\Delta\chi^2=14$.

Two additional features are of note in fig.~\ref{fig:GF chisq}.
First, at $\eps=-1/3$ the $\chi^2$ increases by more than 200 due to an MSW resonance appearing in the neutrino channel at $\sim700$ GeV.
Next at $\eps=-1/12$ we see a sharp feature wherein the $\chi^2$ nearly returns to its value within the standard $3\nu$ scheme.
This is because $\eps=-1/12$ means that the sterile neutrino feels the same NC interaction that the active neutrinos do, and the only difference is due to the effect from vacuum oscillations which are relatively small at these large energies.
While we were preparing this paper, Ref.~\cite{Esmaili:2018qzu} appeared which shows that for $E_\nu<500$ GeV, NSI worsens the agreement with IceCube DeepCore atmospheric neutrino data.
{We note that while it would be possible to evade even the DeepCore bounds with very large NSI $\eps\gtrsim20$ which pushes the resonance below 10 GeV, at that point tight constraints from Super-KamiokaNDE \cite{Mitsuka:2011ty} apply.}

\subsection{The Long-Baseline Constraint from MINOS and MINOS+\label{MINOSbound}}
MINOS and MINOS+ also set strong constraints on the LSND and MiniBooNE best fit 3+1 point. These bounds are dominated by  the MINOS+ CC analysis \cite{Adamson:2017uda}.
Using publicly available data and covariance matrices, we construct a $\chi^2$ including neutrinos and anti-neutrinos, CC and NC, far and near detectors, and appearance and disappearance channels for both MINOS and MINOS+.

We find that, to some extent, it is possible to relax the MINOS and MINOS+ constraint on the 3+1 by including non-zero $\eps$.
The best fit 3+1 point is disfavored relative to the SM at $\Delta\chi^2=24.9$. Within the $3+1+U(1)$ scheme, 
this can be slightly improved relative to the $3+1$ case by $\Delta\chi^2=1.9$, although it is still disfavored compared to the SM at $\Delta\chi^2=22.9$.
The best fit value of NSI is $\eps=0.7^{+0.4}_{-0.5}$.
See the right panel of fig.~\ref{fig:2D GF}.

\subsection{Complete Oscillation Picture\label{ComPic}}
Setting $\theta_{34}=0$, in section \ref{MINOSbound}, we found that
while the IceCube 3+1 picture for $501~{\rm GeV}<E_\nu<10$ TeV is completely ameliorated by the addition of NSI, the MINOS and MINOS+ picture  is only slightly improved.
Moreover, the values of $\eps$ for each turn out to be  quite distinct, with only moderate gains compared to the $3+1$ picture when both experiments are considered simultaneously.
We now check how the pictures change by allowing $\theta_{34}$ to vary.

\begin{figure}
\centering
\includegraphics[width=0.49\textwidth]{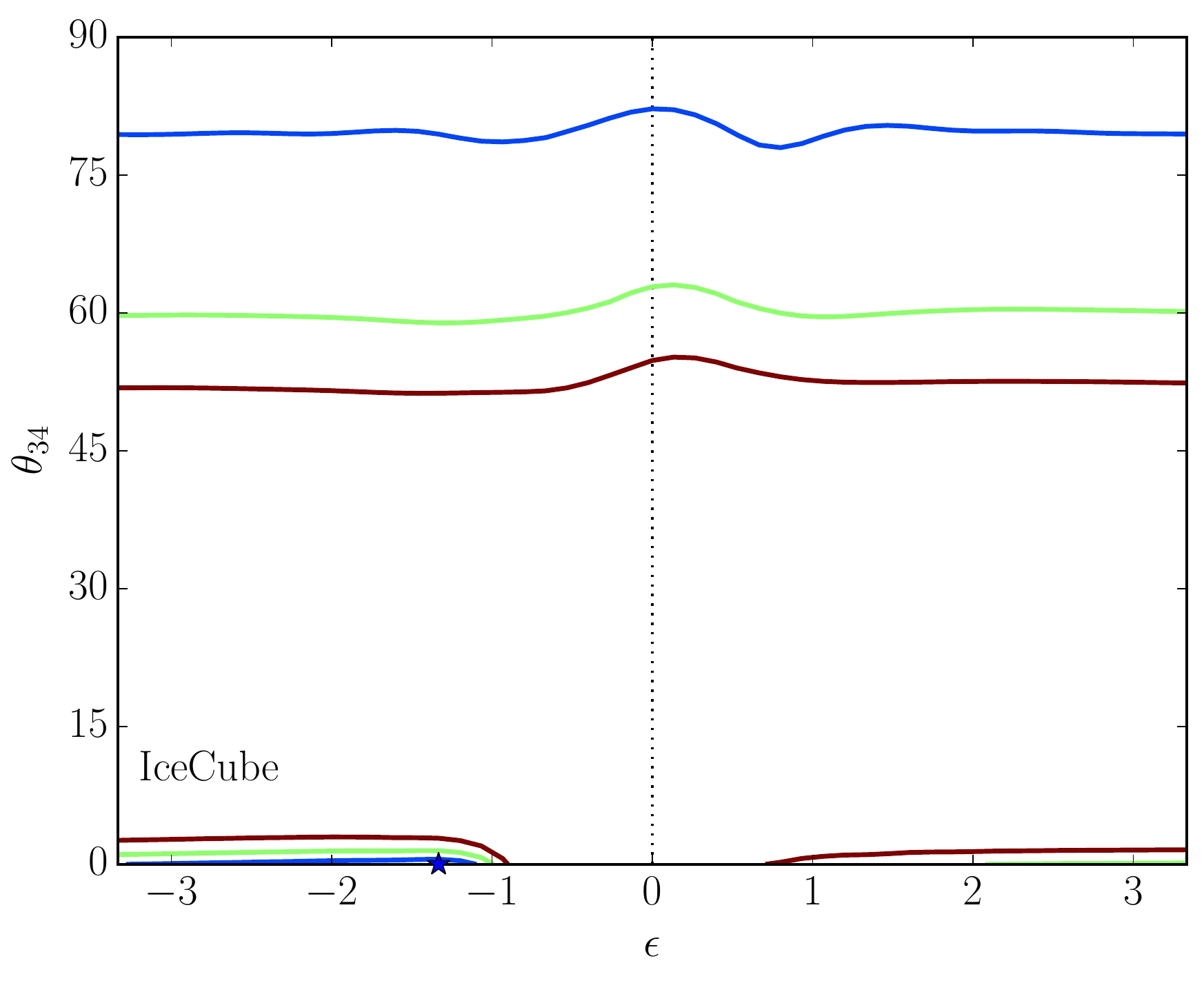}
\includegraphics[width=0.49\textwidth]{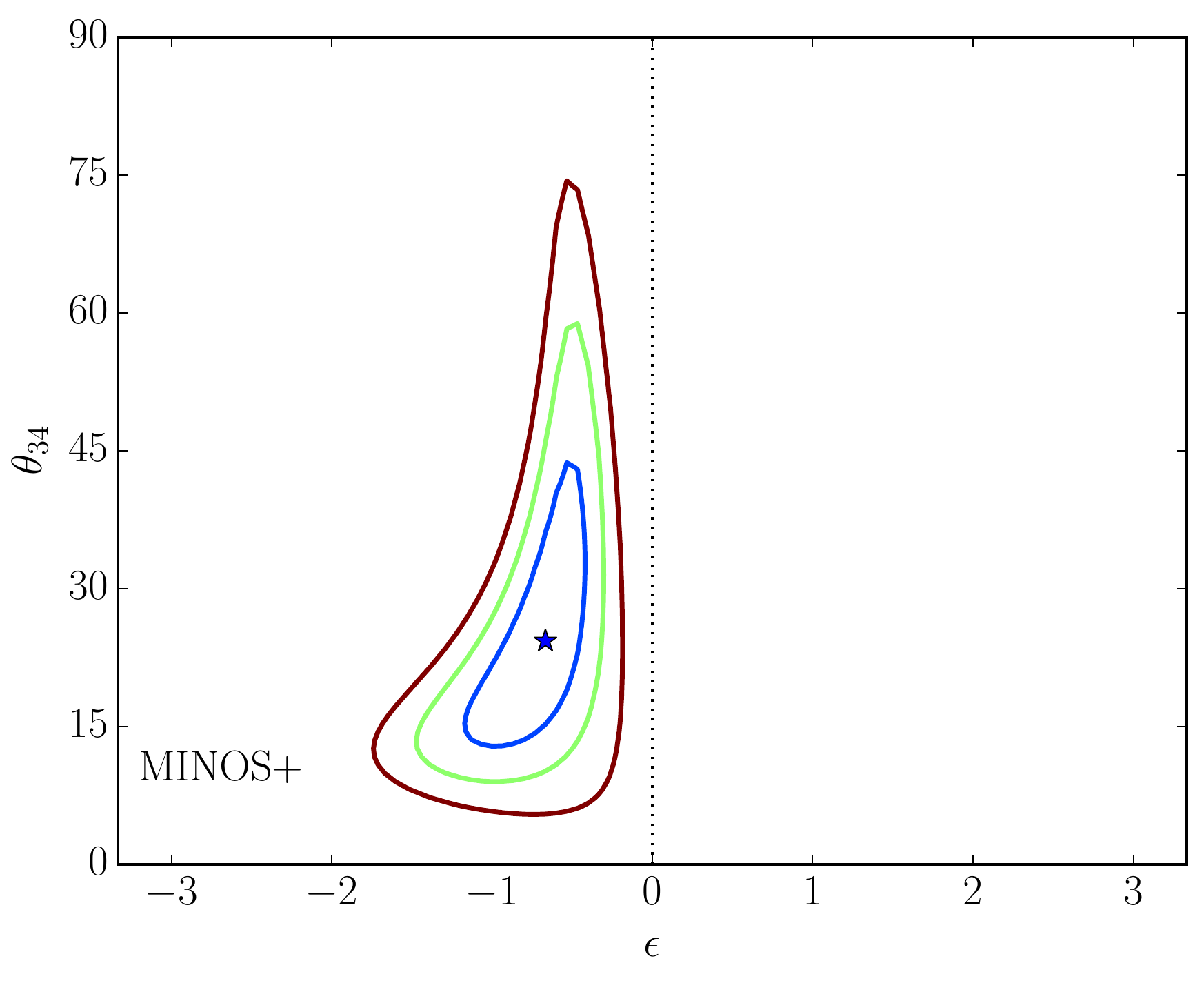}
\caption{The 1, 2, and 3 $\sigma$ best fit regions (blue, green, and red respectively) at 2 d.o.f.~in the $\epsilon$, $\theta_{34}$ space with the best fit point denoted by the star for IceCube (left) and MINOS and MINOS+ (right).
While the best fit point for IceCube provides a good fit to the data and is slightly preferred over the SM, the best fit point for MINOS and MINOS+ is still disfavored compared to the SM by $\Delta\chi^2=10.2$.}
\label{fig:2D GF}
\end{figure}

We again fix $\theta_{14}$, $\theta_{24}$, and $\Delta m^2_{41}$ as above and scan over $\eps$ and $\theta_{34}$.
We find that the MINOS and MINOS+ picture can be considerably improved, but the agreement is still worse than that in the standard $3\nu$ scheme. The best fit point is disfavored relative to the standard $3\nu$ scheme  at $\Delta\chi^2=10.2$ which is still better than the result without $\theta_{34}$ of $\Delta\chi^2=22.9$.
We perform the same analysis with the IceCube data and find that varying $\theta_{34}$ from zero worsens the fit considerably for most values.
Importantly, we did not find any point in the  $\eps-\theta_{34}$ parameter space  which provides suitable fit   both for the MINOS and MINOS+ data  and for the IceCube data  at the same time.
The 1, 2, and 3 $\sigma$ contours for each case are shown in fig.~\ref{fig:2D GF}.

We note that we have still kept the new CP violating phases fixed to zero and these could provide some additional relaxation, although their effect is expected to be small.
A complete global fit varying all  the sterile and NSI parameters is beyond the scope of the present paper.

\section{Summary and discussion \label{summary}}
In the dawn of neutrino precision era, studying possible effects of non-standard  neutrino interaction with matter fields on long baseline and atmospheric neutrino experiments have received significant attention.
From a theory point of view, it is challenging to build viable electroweak symmetric models that give rise to NSI with effective coupling large enough to have discernible  effects in neutrino experiments. One possibility is to invoke light mediators ({see also~\cite{Nelson:2007yq,Engelhardt:2010dx,Wise:2018rnb} for related work}).
It has been shown in the literature that invoking a $U(1)$ gauge symmetry with a gauge boson $(Z')$ at the MeV-scale, phenomenologically relevant NSI can be obtained. For $10~{\rm keV}<m_{Z'}<$few MeV, the cosmological bounds on the extra relativistic degrees of freedom from the BBN and CMB constrain the maximal effective NSI coupling.
 In this paper, we have explored another mass window: $m_{Z'}\sim 10$ eV where the bounds from cosmology are relaxed and the most stringent bound on the matter field coupling to $Z'$ ($<{\rm few} \times 10^{-11}$) 
 comes from the fifth force searches.
  In this mass range, the strongest bound on neutrino coupling $g_{\alpha \beta} \bar{\nu}_\alpha \gamma^\mu \nu_\beta Z'_\mu$ to $Z'$ come from Kaon decay $K^+ \to e^+ \nu_\alpha Z', \mu^+ \nu_\alpha Z'$ which constrains $(\sum_\alpha |g_{e\alpha}|^2)^{1/2},(\sum_\alpha |g_{\mu \alpha}|^2)^{1/2}\lesssim 10^{-9}$. In other words  for
  $m_{Z'}\sim 10$ eV, all $g_{\alpha \beta}$ except  $g_{\tau \tau}$ are constrained to $\lesssim 10^{-9}$. Even with these stringent bounds, active neutrinos can obtain effective NSI coupling of order of $G_F$ or even larger.
 
In  this paper, we have focused on building a model that provides NSI for sterile neutrinos with matter fields.  We have called this scenario the $3+1+U(1)$ model. The motivation for the model is the fact that by introducing NSI for the fourth neutrino of the $3+1$ solution to LSND and MiniBooNE anomalies, the bound from the high energy IceCube  atmospheric neutrino data can be relaxed (for energies greater than 500 GeV).
Our analysis of the atmospheric IceCube data shows that within the $3+1+U(1)$ model,  not only  the IceCube constraint is relaxed,  the quality of fit is actually improved over the fit of standard $3\nu$ scheme, {however we urge the IceCube collaboration to publish a single sterile neutrino analysis spanning from DeepCore energies up through the highest available energies}.
 
Obviously, to obtain an effective interaction between $\nu_s$ and matter fields, the $Z'$ gauge boson has to couple to both $\nu_s$ and matter fields. In our model, the coupling of $Z'$ to gauge boson comes from gauging the 
$B-L$ symmetry so the contributions from electrons and protons of a medium
to the effective neutrino mass  cancel each other out rendering the matter effects  proportional to the neutron number density and independent of the electron and proton density. 
  
As expected, the $\nu_s$ coupling to a new $Z'$ gauge boson is not restricted by terrestrial experiments. However, the requirement of free streaming of active neutrinos mixed with
$\nu_s$ at recombination from one side and the condition of the decay of $\nu_4$ before that era from another side restrict $g_s$ to be around $3 \times 10^{-4}$.
With this value of $g_s$ and saturating the bound on the coupling to quarks, the four-Fermi effective coupling of $\nu_s$ to matter fields, $\epsilon$, can be much larger than 1 but for $U_{si} \sim 0.1$ ($i=1,2,3$), such large values of NSI are ruled out by neutrino oscillation experiments themselves. An effective coupling of 
$\epsilon \sim 1$ which is favored for reconciling the $3+1$ scheme with IceCube atmospheric neutrino data can be easily obtained with a value of coupling to matter fields five orders of magnitudes below the most stringent upper bound. 
 
The mixing of active neutrinos with $\nu_s$ breaks both the new $U_X(1 )$ and electroweak symmetry so it requires a new scalar field, $\phi$, with a nonzero VEV. The requirement of lightness of $Z'$ then implies $\langle \phi \rangle \sim 30$ keV.
The $U_X(1)$ gauge coupling of SM fermions are too small to populate the new light states  in the early universe but $\nu_s$ and consequently the rest of light new state can be produced
via $\nu_a \to \nu_s$ oscillation when $T>$MeV. To prevent such extra relativistic degrees of freedom in our model, $Z'$ also couples to asymmetric dark matter, $\chi$. The forward scattering of $\nu_s$ off the dark matter background for $T>$MeV induces a large effective mass for $\nu_s$ suppressing the effective active sterile mixing and therefore preventing a new contribution to relativistic degrees of freedom at the BBN era.
At $T\sim 100$ keV (well below active neutrino decoupling from plasma) the active sterile mixing resumes its vacuum value and $\nu_a \to \nu_s$ can take place. The produced
$\nu_s$ can in turn populate other light degrees of freedom like $\phi$ and $Z'$ but they decay back to active neutrinos at the onset of structure formation.
We found that if $Z'$coupling to dark matter is chiral, the self-interaction of dark matter particles with such light $Z'$ will be enhanced by $m_\chi^4/m_{Z'}^4$
violating the present bound. We therefore take the coupling to $\chi$ to be  non-chiral.

We carried out a numerical study to test the ability of such a model in reconciling the conflicting evidence on sterile neutrinos.
The addition of a matter effect for the sterile sector provides no change to the short baseline experiments that provide evidence for the sterile neutrino.
As expected, introducing NSI with an effective coupling of the same order of magnitude as $G_F$  completely removes the  constraint from IceCube  atmospheric neutrino data with $500~{\rm GeV}<E_\nu<10$ TeV by returning the oscillation probability to one as it is in the Standard Model.
The next most important constraint on sterile neutrinos comes from MINOS and MINOS+.
We found that including NSI and turning on all three new mixing angles does provide moderate improvement to the quality of fit from the 3+1 case, it is never as good a fit to the data as the standard $3\nu$ scheme.
We examined whether nonzero $\theta_{34}$ can help to relax the bound and found that  the value of $\theta_{34}$ necessary for relaxing the MINOS and MINOS+ bounds is not compatible with that from IceCube.
Ref.~\cite{Esmaili:2018qzu} which appeared when this paper approached conclusion shows that including atmospheric neutrino data with $E_\nu<500$ GeV from IceCube DeepCore \cite{Aartsen:2014yll} disfavors the $3+1+U(1)$ solution to LSND.
We  eagerly anticipate more data from IceCube as their analysis progresses to draw conclusive verdict on the $3+1+U(1) $ solution to the LSND and MiniBooNE anomalies.

Recently \cite{Liao:2018mbg} has shown that reconciling MINOS+ bounds with the $3+1$ solution of MiniBooNE and LSND requires non-standard charged current interaction of $\nu_\mu$ and $\mu$ with quarks with an effective coupling equal to  0.03$G_F$. Such an effective coupling requires a charged mediator on which LHC has set strong bound. For example, if the mediator is a sequential $W$ ($W'$), the bound from the LHC implies that the non-standard charged current effective coupling cannot be larger than  $2 \times 10^{-4}$.  One should also bear in mind that the bounds from MINOS+ have been put into question by \cite{Louis:2018yeg}.

The next generation short baseline experiment, SBN, is being designed to test MiniBooNE results \cite{Antonello:2015lea}. At such short baselines SBN will not be sensitive to 
matter effects so the signature of our $3+1+U(1)$ scenario will be similar to the $3+1$ scheme.
If SBN finds null results and the MiniBooNE anomaly is proved to be a systematic error (for example, due to underestimate of $\pi^0$ events), an upper bound on $|U_{e4}|^2|U_{\mu 4}|^2$ will be set but
the existence 
of $\nu_s$ mixed with the active neutrinos will still remain an intriguing possibility which may be the source of other observed phenomena. Thus, the possibility of activating this fourth neutrinos via the $3+1+U(1)$ scenario introduced in the present paper can be noteworthy independent of MiniBooNE and LSND data.
For example, the ANITA experiment has observed {two upward-pointing events with energies in the EeV$\sim 10^{18}$ eV range \cite{Gorham:2016zah,Gorham:2018ydl}. According to Ref.~\cite{Cherry:2018rxj}, these events can be interpreted as a sign for oscillation of active neutrinos to sterile neutrinos.} The additional matter effects felt by a sterile neutrino in our $3+1+U(1)$ scenario could have interesting implications for these ANITA events, a question we defer to future work.

\acknowledgments
YF is grateful to K Moharrami for useful discussions.
She has received partial funding from the European Union\'~\!s Horizon 2020 research and innovation programme under the Marie Sklodowska-Curie grant agreement No 674896 and No 690575 for this project.
She is also grateful to ICTP associate office for partial financial support.
PBD acknowledges support from the Villum Foundation (Project No.~13164), the Danish National Research Foundation (DNRF91 and Grant No.~1041811001), and the United
States Department of Energy under Grant Contract desc0012704.
IMS is supported by the U.S. Department of Energy under the award number DE-SC0019163.

\bibliographystyle{JHEP}

\bibliography{nu2}

\providecommand{\href}[2]{#2}\begingroup\raggedright\begin{thebibliography}{10}

\bibitem{Aguilar:2001ty}
{\scshape LSND} collaboration, A.~Aguilar-Arevalo et~al., \emph{{Evidence for
  neutrino oscillations from the observation of anti-neutrino(electron)
  appearance in a anti-neutrino(muon) beam}},
  \href{http://dx.doi.org/10.1103/PhysRevD.64.112007}{\emph{Phys. Rev.} {\bf
  D64} (2001) 112007}, [\href{http://arxiv.org/abs/hep-ex/0104049}{{\tt
  hep-ex/0104049}}].

\bibitem{Aguilar-Arevalo:2018gpe}
{\scshape MiniBooNE} collaboration, A.~A. Aguilar-Arevalo et~al.,
  \emph{{Observation of a Significant Excess of Electron-Like Events in the
  MiniBooNE Short-Baseline Neutrino Experiment}},
  \href{http://arxiv.org/abs/1805.12028}{{\tt 1805.12028}}.

\bibitem{Mention:2011rk}
G.~Mention, M.~Fechner, T.~Lasserre, T.~A. Mueller, D.~Lhuillier, M.~Cribier
  et~al., \emph{{The Reactor Antineutrino Anomaly}},
  \href{http://dx.doi.org/10.1103/PhysRevD.83.073006}{\emph{Phys. Rev.} {\bf
  D83} (2011) 073006}, [\href{http://arxiv.org/abs/1101.2755}{{\tt
  1101.2755}}].

\bibitem{Giunti:2010zu}
C.~Giunti and M.~Laveder, \emph{{Statistical Significance of the Gallium
  Anomaly}}, \href{http://dx.doi.org/10.1103/PhysRevC.83.065504}{\emph{Phys.
  Rev.} {\bf C83} (2011) 065504}, [\href{http://arxiv.org/abs/1006.3244}{{\tt
  1006.3244}}].

\bibitem{Nunokawa:2003ep}
H.~Nunokawa, O.~L.~G. Peres and R.~Zukanovich~Funchal, \emph{{Probing the LSND
  mass scale and four neutrino scenarios with a neutrino telescope}},
  \href{http://dx.doi.org/10.1016/S0370-2693(03)00603-8}{\emph{Phys. Lett.}
  {\bf B562} (2003) 279--290}, [\href{http://arxiv.org/abs/hep-ph/0302039}{{\tt
  hep-ph/0302039}}].

\bibitem{Choubey:2007ji}
S.~Choubey, \emph{{Signature of sterile species in atmospheric neutrino data at
  neutrino telescopes}},
  \href{http://dx.doi.org/10.1088/1126-6708/2007/12/014}{\emph{JHEP} {\bf 12}
  (2007) 014}, [\href{http://arxiv.org/abs/0709.1937}{{\tt 0709.1937}}].

\bibitem{Razzaque:2011ab}
S.~Razzaque and A.~{\relax Yu}. Smirnov, \emph{{Searching for sterile neutrinos
  in ice}}, \href{http://dx.doi.org/10.1007/JHEP07(2011)084}{\emph{JHEP} {\bf
  07} (2011) 084}, [\href{http://arxiv.org/abs/1104.1390}{{\tt 1104.1390}}].

\bibitem{Barger:2011rc}
V.~Barger, Y.~Gao and D.~Marfatia, \emph{{Is there evidence for sterile
  neutrinos in IceCube data?}},
  \href{http://dx.doi.org/10.1103/PhysRevD.85.011302}{\emph{Phys. Rev.} {\bf
  D85} (2012) 011302}, [\href{http://arxiv.org/abs/1109.5748}{{\tt
  1109.5748}}].

\bibitem{Esmaili:2012nz}
A.~Esmaili, F.~Halzen and O.~L.~G. Peres, \emph{{Constraining Sterile Neutrinos
  with AMANDA and IceCube Atmospheric Neutrino Data}},
  \href{http://dx.doi.org/10.1088/1475-7516/2012/11/041}{\emph{JCAP} {\bf 1211}
  (2012) 041}, [\href{http://arxiv.org/abs/1206.6903}{{\tt 1206.6903}}].

\bibitem{Esmaili:2013cja}
A.~Esmaili, F.~Halzen and O.~L.~G. Peres, \emph{{Exploring $\nu_\tau - \nu_s$
  mixing with cascade events in DeepCore}},
  \href{http://dx.doi.org/10.1088/1475-7516/2013/07/048}{\emph{JCAP} {\bf 1307}
  (2013) 048}, [\href{http://arxiv.org/abs/1303.3294}{{\tt 1303.3294}}].

\bibitem{Lindner:2015iaa}
M.~Lindner, W.~Rodejohann and X.-J. Xu, \emph{{Sterile neutrinos in the light
  of IceCube}}, \href{http://dx.doi.org/10.1007/JHEP01(2016)124}{\emph{JHEP}
  {\bf 01} (2016) 124}, [\href{http://arxiv.org/abs/1510.00666}{{\tt
  1510.00666}}].

\bibitem{Esmaili:2013vza}
A.~Esmaili and A.~{\relax Yu}. Smirnov, \emph{{Restricting the LSND and
  MiniBooNE sterile neutrinos with the IceCube atmospheric neutrino data}},
  \href{http://dx.doi.org/10.1007/JHEP12(2013)014}{\emph{JHEP} {\bf 12} (2013)
  014}, [\href{http://arxiv.org/abs/1307.6824}{{\tt 1307.6824}}].

\bibitem{TheIceCube:2016oqi}
{\scshape IceCube} collaboration, M.~G. Aartsen et~al., \emph{{Searches for
  Sterile Neutrinos with the IceCube Detector}},
  \href{http://dx.doi.org/10.1103/PhysRevLett.117.071801}{\emph{Phys. Rev.
  Lett.} {\bf 117} (2016) 071801}, [\href{http://arxiv.org/abs/1605.01990}{{\tt
  1605.01990}}].

\bibitem{Liao:2016reh}
J.~Liao and D.~Marfatia, \emph{{Impact of nonstandard interactions on sterile
  neutrino searches at IceCube}},
  \href{http://dx.doi.org/10.1103/PhysRevLett.117.071802}{\emph{Phys. Rev.
  Lett.} {\bf 117} (2016) 071802}, [\href{http://arxiv.org/abs/1602.08766}{{\tt
  1602.08766}}].

\bibitem{huang_en_chuan_2018_1287004}
E.-C. Huang, \emph{Updated miniboone $\nu_\mu\to\nu_e$ oscillation on results},
   June, 2018.
\newblock 10.5281/zenodo.1287004.

\bibitem{Capozzi:2017auw}
F.~Capozzi, I.~M. Shoemaker and L.~Vecchi, \emph{{Solar Neutrinos as a Probe of
  Dark Matter-Neutrino Interactions}},
  \href{http://dx.doi.org/10.1088/1475-7516/2017/07/021}{\emph{JCAP} {\bf 1707}
  (2017) 021}, [\href{http://arxiv.org/abs/1702.08464}{{\tt 1702.08464}}].

\bibitem{Hannestad:2013ana}
S.~Hannestad, R.~S. Hansen and T.~Tram, \emph{{How Self-Interactions can
  Reconcile Sterile Neutrinos with Cosmology}},
  \href{http://dx.doi.org/10.1103/PhysRevLett.112.031802}{\emph{Phys. Rev.
  Lett.} {\bf 112} (2014) 031802}, [\href{http://arxiv.org/abs/1310.5926}{{\tt
  1310.5926}}].

\bibitem{Dasgupta:2013zpn}
B.~Dasgupta and J.~Kopp, \emph{{Cosmologically Safe eV-Scale Sterile Neutrinos
  and Improved Dark Matter Structure}},
  \href{http://dx.doi.org/10.1103/PhysRevLett.112.031803}{\emph{Phys. Rev.
  Lett.} {\bf 112} (2014) 031803}, [\href{http://arxiv.org/abs/1310.6337}{{\tt
  1310.6337}}].

\bibitem{Mirizzi:2014ama}
A.~Mirizzi, G.~Mangano, O.~Pisanti and N.~Saviano, \emph{{Collisional
  production of sterile neutrinos via secret interactions and cosmological
  implications}},
  \href{http://dx.doi.org/10.1103/PhysRevD.91.025019}{\emph{Phys. Rev.} {\bf
  D91} (2015) 025019}, [\href{http://arxiv.org/abs/1410.1385}{{\tt
  1410.1385}}].

\bibitem{Cherry:2014xra}
J.~F. Cherry, A.~Friedland and I.~M. Shoemaker, \emph{{Neutrino Portal Dark
  Matter: From Dwarf Galaxies to IceCube}},
  \href{http://arxiv.org/abs/1411.1071}{{\tt 1411.1071}}.

\bibitem{Chu:2015ipa}
X.~Chu, B.~Dasgupta and
  J.~Kopp\href{http://dx.doi.org/10.1088/1475-7516/2015/10/011}{\emph{JCAP}
  {\bf 1510} (2015) 011}, [\href{http://arxiv.org/abs/1505.02795}{{\tt
  1505.02795}}].

\bibitem{Cherry:2016jol}
J.~F. Cherry, A.~Friedland and I.~M. Shoemaker, \emph{{Short-baseline neutrino
  oscillations, Planck, and IceCube}},
  \href{http://arxiv.org/abs/1605.06506}{{\tt 1605.06506}}.

\bibitem{Vecchi:2016lty}
L.~Vecchi, \emph{{Light sterile neutrinos from a late phase transition}},
  \href{http://dx.doi.org/10.1103/PhysRevD.94.113015}{\emph{Phys. Rev.} {\bf
  D94} (2016) 113015}, [\href{http://arxiv.org/abs/1607.04161}{{\tt
  1607.04161}}].

\bibitem{Saviano:2014esa}
N.~Saviano, O.~Pisanti, G.~Mangano and A.~Mirizzi, \emph{{Unveiling secret
  interactions among sterile neutrinos with big-bang nucleosynthesis}},
  \href{http://dx.doi.org/10.1103/PhysRevD.90.113009}{\emph{Phys. Rev.} {\bf
  D90} (2014) 113009}, [\href{http://arxiv.org/abs/1409.1680}{{\tt
  1409.1680}}].

\bibitem{Ade:2015xua}
{\scshape Planck} collaboration, P.~A.~R. Ade et~al., \emph{{Planck 2015
  results. XIII. Cosmological parameters}},
  \href{http://dx.doi.org/10.1051/0004-6361/201525830}{\emph{Astron.
  Astrophys.} {\bf 594} (2016) A13},
  [\href{http://arxiv.org/abs/1502.01589}{{\tt 1502.01589}}].

\bibitem{Song:2018zyl}
N.~Song, M.~C. Gonzalez-Garcia and J.~Salvado, \emph{{Cosmological constraints
  with self-interacting sterile neutrinos}},
  \href{http://arxiv.org/abs/1805.08218}{{\tt 1805.08218}}.

\bibitem{Adamson:2017zcg}
{\scshape NOvA} collaboration, P.~Adamson et~al., \emph{{Search for
  active-sterile neutrino mixing using neutral-current interactions in NOvA}},
  \href{http://dx.doi.org/10.1103/PhysRevD.96.072006}{\emph{Phys. Rev.} {\bf
  D96} (2017) 072006}, [\href{http://arxiv.org/abs/1706.04592}{{\tt
  1706.04592}}].

\bibitem{Adamson:2017uda}
{\scshape MINOS} collaboration, P.~Adamson et~al., \emph{{Search for sterile
  neutrinos in MINOS and MINOS+ using a two-detector fit}}, {\emph{Submitted
  to: Phys. Rev. Lett.} (2017) }, [\href{http://arxiv.org/abs/1710.06488}{{\tt
  1710.06488}}].

\bibitem{Louis:2018yeg}
W.~C. Louis, \emph{{Problems With the MINOS/MINOS+ Sterile Neutrino $\nu _\mu$
  Result}},  \href{http://arxiv.org/abs/1803.11488}{{\tt 1803.11488}}.

\bibitem{Redondo:2013lna}
J.~Redondo and G.~Raffelt, \emph{{Solar constraints on hidden photons
  re-visited}},
  \href{http://dx.doi.org/10.1088/1475-7516/2013/08/034}{\emph{JCAP} {\bf 1308}
  (2013) 034}, [\href{http://arxiv.org/abs/1305.2920}{{\tt 1305.2920}}].

\bibitem{Hardy:2016kme}
E.~Hardy and R.~Lasenby, \emph{{Stellar cooling bounds on new light particles:
  plasma mixing effects}},
  \href{http://dx.doi.org/10.1007/JHEP02(2017)033}{\emph{JHEP} {\bf 02} (2017)
  033}, [\href{http://arxiv.org/abs/1611.05852}{{\tt 1611.05852}}].

\bibitem{Leeb:1992qf}
H.~Leeb and J.~Schmiedmayer, \emph{{Constraint on hypothetical light
  interacting bosons from low-energy neutron experiments}},
  \href{http://dx.doi.org/10.1103/PhysRevLett.68.1472}{\emph{Phys. Rev. Lett.}
  {\bf 68} (1992) 1472--1475}.

\bibitem{Artamonov:2009sz}
{\scshape BNL-E949} collaboration, A.~V. Artamonov et~al., \emph{{Study of the
  decay $K^+\to\pi^+\nu \bar\nu$ in the momentum region $140 < P_\pi < 199$
  MeV/c}}, \href{http://dx.doi.org/10.1103/PhysRevD.79.092004}{\emph{Phys.
  Rev.} {\bf D79} (2009) 092004}, [\href{http://arxiv.org/abs/0903.0030}{{\tt
  0903.0030}}].

\bibitem{Batell:2014yra}
B.~Batell, P.~deNiverville, D.~McKeen, M.~Pospelov and A.~Ritz,
  \emph{{Leptophobic Dark Matter at Neutrino Factories}},
  \href{http://dx.doi.org/10.1103/PhysRevD.90.115014}{\emph{Phys. Rev.} {\bf
  D90} (2014) 115014}, [\href{http://arxiv.org/abs/1405.7049}{{\tt
  1405.7049}}].

\bibitem{Bordag:2001qi}
M.~Bordag, U.~Mohideen and V.~M. Mostepanenko, \emph{{New developments in the
  Casimir effect}},
  \href{http://dx.doi.org/10.1016/S0370-1573(01)00015-1}{\emph{Phys. Rept.}
  {\bf 353} (2001) 1--205}, [\href{http://arxiv.org/abs/quant-ph/0106045}{{\tt
  quant-ph/0106045}}].

\bibitem{Gninenko:1998pm}
S.~N. Gninenko and N.~V. Krasnikov, \emph{{On search for a new light gauge
  boson from pi0 (eta) $\to$ gamma + X decays in neutrino experiments}},
  \href{http://dx.doi.org/10.1016/S0370-2693(98)00358-X}{\emph{Phys. Lett.}
  {\bf B427} (1998) 307--313}, [\href{http://arxiv.org/abs/hep-ph/9802375}{{\tt
  hep-ph/9802375}}].

\bibitem{Altegoer:1998qta}
{\scshape NOMAD} collaboration, J.~Altegoer et~al., \emph{{Search for a new
  gauge boson in pi0 decays}},
  \href{http://dx.doi.org/10.1016/S0370-2693(98)00402-X}{\emph{Phys. Lett.}
  {\bf B428} (1998) 197--205}, [\href{http://arxiv.org/abs/hep-ex/9804003}{{\tt
  hep-ex/9804003}}].

\bibitem{Bakhti:2017jhm}
P.~Bakhti and Y.~Farzan, \emph{{Constraining secret gauge interactions of
  neutrinos by meson decays}},
  \href{http://dx.doi.org/10.1103/PhysRevD.95.095008}{\emph{Phys. Rev.} {\bf
  D95} (2017) 095008}, [\href{http://arxiv.org/abs/1702.04187}{{\tt
  1702.04187}}].

\bibitem{Laha:2013xua}
R.~Laha, B.~Dasgupta and J.~F. Beacom, \emph{{Constraints on New Neutrino
  Interactions via Light Abelian Vector Bosons}},
  \href{http://dx.doi.org/10.1103/PhysRevD.89.093025}{\emph{Phys. Rev.} {\bf
  D89} (2014) 093025}, [\href{http://arxiv.org/abs/1304.3460}{{\tt
  1304.3460}}].

\bibitem{Bakhti:2018avv}
P.~Bakhti, Y.~Farzan and M.~Rajaee, \emph{{Secret interactions of neutrinos
  with light gauge boson at the DUNE near detector}},
  \href{http://arxiv.org/abs/1810.04441}{{\tt 1810.04441}}.

\bibitem{Archidiacono:2015oma}
M.~Archidiacono, S.~Hannestad, R.~S. Hansen and T.~Tram, \emph{{Sterile
  neutrinos with pseudoscalar self-interactions and cosmology}},
  \href{http://dx.doi.org/10.1103/PhysRevD.93.045004}{\emph{Phys. Rev.} {\bf
  D93} (2016) 045004}, [\href{http://arxiv.org/abs/1508.02504}{{\tt
  1508.02504}}].

\bibitem{Kopp:2014fha}
J.~Kopp and J.~Welter, \emph{{The Not-So-Sterile 4th Neutrino: Constraints on
  New Gauge Interactions from Neutrino Oscillation Experiments}},
  \href{http://dx.doi.org/10.1007/JHEP12(2014)104}{\emph{JHEP} {\bf 12} (2014)
  104}, [\href{http://arxiv.org/abs/1408.0289}{{\tt 1408.0289}}].

\bibitem{Kaplan:1991ah}
D.~B. Kaplan, \emph{{A Single explanation for both the baryon and dark matter
  densities}}, \href{http://dx.doi.org/10.1103/PhysRevLett.68.741}{\emph{Phys.
  Rev. Lett.} {\bf 68} (1992) 741--743}.

\bibitem{Nussinov:1985xr}
S.~Nussinov, \emph{{TECHNOCOSMOLOGY: COULD A TECHNIBARYON EXCESS PROVIDE A
  'NATURAL' MISSING MASS CANDIDATE?}},
  \href{http://dx.doi.org/10.1016/0370-2693(85)90689-6}{\emph{Phys. Lett.} {\bf
  165B} (1985) 55--58}.

\bibitem{Barr:1991qn}
S.~M. Barr, \emph{{Baryogenesis, sphalerons and the cogeneration of dark
  matter}}, \href{http://dx.doi.org/10.1103/PhysRevD.44.3062}{\emph{Phys. Rev.}
  {\bf D44} (1991) 3062--3066}.

\bibitem{Barr:1990ca}
S.~M. Barr, R.~S. Chivukula and E.~Farhi, \emph{{Electroweak Fermion Number
  Violation and the Production of Stable Particles in the Early Universe}},
  \href{http://dx.doi.org/10.1016/0370-2693(90)91661-T}{\emph{Phys. Lett.} {\bf
  B241} (1990) 387--391}.

\bibitem{Gudnason:2006ug}
S.~B. Gudnason, C.~Kouvaris and F.~Sannino, \emph{{Towards working technicolor:
  Effective theories and dark matter}},
  \href{http://dx.doi.org/10.1103/PhysRevD.73.115003}{\emph{Phys. Rev.} {\bf
  D73} (2006) 115003}, [\href{http://arxiv.org/abs/hep-ph/0603014}{{\tt
  hep-ph/0603014}}].

\bibitem{Dodelson:1991iv}
S.~Dodelson, B.~R. Greene and L.~M. Widrow, \emph{{Baryogenesis, dark matter
  and the width of the Z}},
  \href{http://dx.doi.org/10.1016/0550-3213(92)90328-9}{\emph{Nucl. Phys.} {\bf
  B372} (1992) 467--493}.

\bibitem{Fujii:2002aj}
M.~Fujii and T.~Yanagida, \emph{{A Solution to the coincidence puzzle of
  Omega(B) and Omega (DM)}},
  \href{http://dx.doi.org/10.1016/S0370-2693(02)02341-9}{\emph{Phys. Lett.}
  {\bf B542} (2002) 80--88}, [\href{http://arxiv.org/abs/hep-ph/0206066}{{\tt
  hep-ph/0206066}}].

\bibitem{Kitano:2004sv}
R.~Kitano and I.~Low, \emph{{Dark matter from baryon asymmetry}},
  \href{http://dx.doi.org/10.1103/PhysRevD.71.023510}{\emph{Phys. Rev.} {\bf
  D71} (2005) 023510}, [\href{http://arxiv.org/abs/hep-ph/0411133}{{\tt
  hep-ph/0411133}}].

\bibitem{Kitano:2008tk}
R.~Kitano, H.~Murayama and M.~Ratz, \emph{{Unified origin of baryons and dark
  matter}}, \href{http://dx.doi.org/10.1016/j.physletb.2008.09.049}{\emph{Phys.
  Lett.} {\bf B669} (2008) 145--149},
  [\href{http://arxiv.org/abs/0807.4313}{{\tt 0807.4313}}].

\bibitem{Farrar:2005zd}
G.~R. Farrar and G.~Zaharijas, \emph{{Dark matter and the baryon asymmetry}},
  \href{http://dx.doi.org/10.1103/PhysRevLett.96.041302}{\emph{Phys. Rev.
  Lett.} {\bf 96} (2006) 041302},
  [\href{http://arxiv.org/abs/hep-ph/0510079}{{\tt hep-ph/0510079}}].

\bibitem{Kaplan:2009ag}
D.~E. Kaplan, M.~A. Luty and K.~M. Zurek, \emph{{Asymmetric Dark Matter}},
  \href{http://dx.doi.org/10.1103/PhysRevD.79.115016}{\emph{Phys. Rev.} {\bf
  D79} (2009) 115016}, [\href{http://arxiv.org/abs/0901.4117}{{\tt
  0901.4117}}].

\bibitem{Tulin:2013teo}
S.~Tulin, H.-B. Yu and K.~M. Zurek, \emph{{Beyond Collisionless Dark Matter:
  Particle Physics Dynamics for Dark Matter Halo Structure}},
  \href{http://dx.doi.org/10.1103/PhysRevD.87.115007}{\emph{Phys. Rev.} {\bf
  D87} (2013) 115007}, [\href{http://arxiv.org/abs/1302.3898}{{\tt
  1302.3898}}].

\bibitem{Clowe:2006eq}
D.~Clowe, M.~Bradac, A.~H. Gonzalez, M.~Markevitch, S.~W. Randall, C.~Jones
  et~al., \emph{{A direct empirical proof of the existence of dark matter}},
  \href{http://dx.doi.org/10.1086/508162}{\emph{Astrophys. J.} {\bf 648} (2006)
  L109--L113}, [\href{http://arxiv.org/abs/astro-ph/0608407}{{\tt
  astro-ph/0608407}}].

\bibitem{Randall:2007ph}
S.~W. Randall, M.~Markevitch, D.~Clowe, A.~H. Gonzalez and M.~Bradac,
  \emph{{Constraints on the Self-Interaction Cross-Section of Dark Matter from
  Numerical Simulations of the Merging Galaxy Cluster 1E 0657-56}},
  \href{http://dx.doi.org/10.1086/587859}{\emph{Astrophys. J.} {\bf 679} (2008)
  1173--1180}, [\href{http://arxiv.org/abs/0704.0261}{{\tt 0704.0261}}].

\bibitem{Ackerman:mha}
L.~Ackerman, M.~R. Buckley, S.~M. Carroll and M.~Kamionkowski, \emph{{Dark
  Matter and Dark Radiation}},
  \href{http://dx.doi.org/10.1103/PhysRevD.79.023519,
  10.1142/9789814293792_0021}{\emph{Phys. Rev.} {\bf D79} (2009) 023519},
  [\href{http://arxiv.org/abs/0810.5126}{{\tt 0810.5126}}].

\bibitem{Karagiorgi:2012kw}
G.~Karagiorgi, M.~H. Shaevitz and J.~M. Conrad, \emph{{Confronting the
  Short-Baseline Oscillation Anomalies with a Single Sterile Neutrino and
  Non-Standard Matter Effects}},  \href{http://arxiv.org/abs/1202.1024}{{\tt
  1202.1024}}.

\bibitem{Farzan:2016wym}
Y.~Farzan and J.~Heeck, \emph{{Neutrinophilic nonstandard interactions}},
  \href{http://dx.doi.org/10.1103/PhysRevD.94.053010}{\emph{Phys. Rev.} {\bf
  D94} (2016) 053010}, [\href{http://arxiv.org/abs/1607.07616}{{\tt
  1607.07616}}].

\bibitem{Patwardhan:2014iha}
A.~V. Patwardhan and G.~M. Fuller, \emph{{Late-time vacuum phase transitions:
  Connecting sub-eV scale physics with cosmological structure formation}},
  \href{http://dx.doi.org/10.1103/PhysRevD.90.063009}{\emph{Phys. Rev.} {\bf
  D90} (2014) 063009}, [\href{http://arxiv.org/abs/1401.1923}{{\tt
  1401.1923}}].

\bibitem{Pontecorvo:1967fh}
B.~Pontecorvo, \emph{{Neutrino Experiments and the Problem of Conservation of
  Leptonic Charge}}, {\emph{Sov. Phys. JETP} {\bf 26} (1968) 984--988}.

\bibitem{Maki:1962mu}
Z.~Maki, M.~Nakagawa and S.~Sakata, \emph{{Remarks on the unified model of
  elementary particles}},
  \href{http://dx.doi.org/10.1143/PTP.28.870}{\emph{Prog. Theor. Phys.} {\bf
  28} (1962) 870--880}.

\bibitem{Aartsen:2017bap}
{\scshape IceCube} collaboration, M.~G. Aartsen et~al., \emph{{Search for
  sterile neutrino mixing using three years of IceCube DeepCore data}},
  \href{http://dx.doi.org/10.1103/PhysRevD.95.112002}{\emph{Phys. Rev.} {\bf
  D95} (2017) 112002}, [\href{http://arxiv.org/abs/1702.05160}{{\tt
  1702.05160}}].

\bibitem{Dziewonski:1981xy}
A.~M. Dziewonski and D.~L. Anderson, \emph{{Preliminary reference earth
  model}}, \href{http://dx.doi.org/10.1016/0031-9201(81)90046-7}{\emph{Phys.
  Earth Planet. Interiors} {\bf 25} (1981) 297--356}.

\bibitem{Esteban:2016qun}
I.~Esteban, M.~C. Gonzalez-Garcia, M.~Maltoni, I.~Martinez-Soler and
  T.~Schwetz, \emph{{Updated fit to three neutrino mixing: exploring the
  accelerator-reactor complementarity}},
  \href{http://dx.doi.org/10.1007/JHEP01(2017)087}{\emph{JHEP} {\bf 01} (2017)
  087}, [\href{http://arxiv.org/abs/1611.01514}{{\tt 1611.01514}}].

\bibitem{nu-fit:v3.2}
``{NuFIT 3.2}.'' \href{http://www.nu-fit.org}{nu-fit.org}, 2018.

\bibitem{Dentler:2018sju}
M.~Dentler, A.~Hernandez-Cabezudo, J.~Kopp, P.~A.~N. Machado, M.~Maltoni,
  I.~Martinez-Soler et~al., \emph{{Updated Global Analysis of Neutrino
  Oscillations in the Presence of eV-Scale Sterile Neutrinos}},
  \href{http://dx.doi.org/10.1007/JHEP08(2018)010}{\emph{JHEP} {\bf 08} (2018)
  010}, [\href{http://arxiv.org/abs/1803.10661}{{\tt 1803.10661}}].

\bibitem{An:2017osx}
{\scshape Daya Bay} collaboration, F.~P. An et~al., \emph{{Evolution of the
  Reactor Antineutrino Flux and Spectrum at Daya Bay}},
  \href{http://dx.doi.org/10.1103/PhysRevLett.118.251801}{\emph{Phys. Rev.
  Lett.} {\bf 118} (2017) 251801}, [\href{http://arxiv.org/abs/1704.01082}{{\tt
  1704.01082}}].

\bibitem{Hayes:2017res}
A.~C. Hayes, G.~Jungman, E.~A. McCutchan, A.~A. Sonzogni, G.~T. Garvey and
  X.~Wang, \emph{{Analysis of the Daya Bay Reactor Antineutrino Flux Changes
  with Fuel Burnup}},
  \href{http://dx.doi.org/10.1103/PhysRevLett.120.022503}{\emph{Phys. Rev.
  Lett.} {\bf 120} (2018) 022503}, [\href{http://arxiv.org/abs/1707.07728}{{\tt
  1707.07728}}].

\bibitem{Wolfenstein:1977ue}
L.~Wolfenstein, \emph{{Neutrino Oscillations in Matter}},
  \href{http://dx.doi.org/10.1103/PhysRevD.17.2369}{\emph{Phys.Rev.} {\bf D17}
  (1978) 2369--2374}.

\bibitem{Mikheev:1986gs}
S.~P. Mikheev and A.~{\relax Yu}. Smirnov, \emph{{Resonance Amplification of
  Oscillations in Matter and Spectroscopy of Solar Neutrinos}}, {\emph{Sov. J.
  Nucl. Phys.} {\bf 42} (1985) 913--917}.

\bibitem{Aartsen:2015rwa}
{\scshape IceCube} collaboration, M.~G. Aartsen et~al., \emph{{Evidence for
  Astrophysical Muon Neutrinos from the Northern Sky with IceCube}},
  \href{http://dx.doi.org/10.1103/PhysRevLett.115.081102}{\emph{Phys. Rev.
  Lett.} {\bf 115} (2015) 081102}, [\href{http://arxiv.org/abs/1507.04005}{{\tt
  1507.04005}}].

\bibitem{IC:HE_NuMu_diffuse}
\url{https://icecube.wisc.edu/science/data/HE_NuMu_diffuse/}.

\bibitem{Honda:2004yz}
M.~Honda, T.~Kajita, K.~Kasahara and S.~Midorikawa, \emph{{A New calculation of
  the atmospheric neutrino flux in a 3-dimensional scheme}},
  \href{http://dx.doi.org/10.1103/PhysRevD.70.043008}{\emph{Phys. Rev.} {\bf
  D70} (2004) 043008}, [\href{http://arxiv.org/abs/astro-ph/0404457}{{\tt
  astro-ph/0404457}}].

\bibitem{Aartsen:2013eka}
{\scshape IceCube} collaboration, M.~G. Aartsen et~al., \emph{{Search for a
  diffuse flux of astrophysical muon neutrinos with the IceCube 59-string
  configuration}},
  \href{http://dx.doi.org/10.1103/PhysRevD.89.062007}{\emph{Phys. Rev.} {\bf
  D89} (2014) 062007}, [\href{http://arxiv.org/abs/1311.7048}{{\tt
  1311.7048}}].

\bibitem{Gaisser:2011cc}
T.~K. Gaisser, \emph{{Spectrum of cosmic-ray nucleons, kaon production, and the
  atmospheric muon charge ratio}},
  \href{http://dx.doi.org/10.1016/j.astropartphys.2012.02.010}{\emph{Astropart.
  Phys.} {\bf 35} (2012) 801--806}, [\href{http://arxiv.org/abs/1111.6675}{{\tt
  1111.6675}}].

\bibitem{Fogli:2002pt}
G.~L. Fogli, E.~Lisi, A.~Marrone, D.~Montanino and A.~Palazzo, \emph{{Getting
  the most from the statistical analysis of solar neutrino oscillations}},
  \href{http://dx.doi.org/10.1103/PhysRevD.66.053010}{\emph{Phys. Rev.} {\bf
  D66} (2002) 053010}, [\href{http://arxiv.org/abs/hep-ph/0206162}{{\tt
  hep-ph/0206162}}].

\bibitem{Esmaili:2018qzu}
A.~Esmaili and H.~Nunokawa, \emph{{On the robustness of IceCube's bound on
  sterile neutrinos in the presence of non-standard interactions}},
  \href{http://arxiv.org/abs/1810.11940}{{\tt 1810.11940}}.

\bibitem{Mitsuka:2011ty}
{\scshape Super-Kamiokande} collaboration, G.~Mitsuka et~al., \emph{{Study of
  Non-Standard Neutrino Interactions with Atmospheric Neutrino Data in
  Super-Kamiokande I and II}},
  \href{http://dx.doi.org/10.1103/PhysRevD.84.113008}{\emph{Phys. Rev.} {\bf
  D84} (2011) 113008}, [\href{http://arxiv.org/abs/1109.1889}{{\tt
  1109.1889}}].

\bibitem{Nelson:2007yq}
A.~E. Nelson and J.~Walsh, \emph{{Short Baseline Neutrino Oscillations and a
  New Light Gauge Boson}},
  \href{http://dx.doi.org/10.1103/PhysRevD.77.033001}{\emph{Phys. Rev.} {\bf
  D77} (2008) 033001}, [\href{http://arxiv.org/abs/0711.1363}{{\tt
  0711.1363}}].

\bibitem{Engelhardt:2010dx}
N.~Engelhardt, A.~E. Nelson and J.~R. Walsh, \emph{{Apparent CPT Violation in
  Neutrino Oscillation Experiments}},
  \href{http://dx.doi.org/10.1103/PhysRevD.81.113001}{\emph{Phys. Rev.} {\bf
  D81} (2010) 113001}, [\href{http://arxiv.org/abs/1002.4452}{{\tt
  1002.4452}}].

\bibitem{Wise:2018rnb}
M.~B. Wise and Y.~Zhang, \emph{{Lepton Flavorful Fifth Force and
  Depth-dependent Neutrino Matter Interactions}},
  \href{http://dx.doi.org/10.1007/JHEP06(2018)053}{\emph{JHEP} {\bf 06} (2018)
  053}, [\href{http://arxiv.org/abs/1803.00591}{{\tt 1803.00591}}].

\bibitem{Aartsen:2014yll}
{\scshape IceCube} collaboration, M.~G. Aartsen et~al., \emph{{Determining
  neutrino oscillation parameters from atmospheric muon neutrino disappearance
  with three years of IceCube DeepCore data}},
  \href{http://dx.doi.org/10.1103/PhysRevD.91.072004}{\emph{Phys. Rev.} {\bf
  D91} (2015) 072004}, [\href{http://arxiv.org/abs/1410.7227}{{\tt
  1410.7227}}].

\bibitem{Liao:2018mbg}
J.~Liao, D.~Marfatia and K.~Whisnant, \emph{{MiniBooNE, MINOS+ and IceCube data
  imply a baroque neutrino sector}},
  \href{http://arxiv.org/abs/1810.01000}{{\tt 1810.01000}}.

\bibitem{Antonello:2015lea}
{\scshape LAr1-ND, ICARUS-WA104, MicroBooNE} collaboration, M.~Antonello
  et~al., \emph{{A Proposal for a Three Detector Short-Baseline Neutrino
  Oscillation Program in the Fermilab Booster Neutrino Beam}},
  \href{http://arxiv.org/abs/1503.01520}{{\tt 1503.01520}}.

\bibitem{Gorham:2016zah}
P.~W. Gorham et~al., \emph{{Characteristics of Four Upward-pointing
  Cosmic-ray-like Events Observed with ANITA}},
  \href{http://dx.doi.org/10.1103/PhysRevLett.117.071101}{\emph{Phys. Rev.
  Lett.} {\bf 117} (2016) 071101}, [\href{http://arxiv.org/abs/1603.05218}{{\tt
  1603.05218}}].

\bibitem{Gorham:2018ydl}
{\scshape ANITA} collaboration, P.~W. Gorham et~al., \emph{{Observation of an
  Unusual Upward-going Cosmic-ray-like Event in the Third Flight of ANITA}},
  \href{http://dx.doi.org/10.1103/PhysRevLett.121.161102}{\emph{Phys. Rev.
  Lett.} {\bf 121} (2018) 161102}, [\href{http://arxiv.org/abs/1803.05088}{{\tt
  1803.05088}}].

\bibitem{Cherry:2018rxj}
J.~F. Cherry and I.~Shoemaker, \emph{{A Sterile Neutrino Origin for the Upward
  Directed Cosmic Ray Shower Detected by ANITA}},
  \href{http://arxiv.org/abs/1802.01611}{{\tt 1802.01611}}.

\end{thebibliography}\endgroup

\end{document}